\expandafter\edef\csname hypers@fe\endcsname{\catcode
                                             `\noexpand @=\the\catcode`\@}%
\catcode`\@=11
%
% Check if the file is already included
%
\ifx\hyperd@ne\hyper@ndefined
 \global\let\hyperd@ne=\relax
\else
 \errhelp{hyperbasics.tex needs to be included only once outside
          of any {...} or \begingroup...\endgroup. You have tried to
          include it more than once. If the previous include was indeed
          outside any groupings, continue and all will be well.}%
 \errmessage{Input this file only once!}%
  
\fi
%
% Version number
%
\def\hyperv@rsion{8}%
%
% Check and input a previous .hrf file if it exists
%
\newread\hyperf@le
\def\hyperf@lename{\jobname.hrf}%
\immediate\openin\hyperf@le\hyperf@lename\relax
\ifeof\hyperf@le\relax
 \immediate\closein\hyperf@le\relax
\else
 \immediate\closein\hyperf@le\relax
 \input \hyperf@lename
\fi
%
% Open a new .hrf file
%
\newwrite\hyperf@le
\immediate\openout\hyperf@le\hyperf@lename
%%%%
% MAIN SECTION
%%%%
%
% define a token register
%
\newtoks\hypert@ks
%
% Define a convenient macro to hold the character #
%
\edef\hypert@mp{\catcode`\noexpand\#=\the\catcode`\#}%
\catcode`\#=12
\def\hyperh@sh{#}%
\hypert@mp
\let\hypert@mp=\relax
\let\hyper@nd=\relax
\def\hyperstr@pquote"#1"#2\hyper@nd{\ifx\hyper@ndefined#2\hyper@ndefined#1\else
                                    \ifx\hyper@ndefined#1\hyper@ndefined
                                    \hyperstr@pquote#2"\hyper@nd\else
                                    #1\hyperstr@pquote"#2"\hyper@nd\fi\fi}%
\def\hyperstr@pblank" #1 #2\hyper@nd"{\ifx\hyper@ndefined#2\hyper@ndefined#1\else
                                    \ifx\hyper@ndefined#1\hyper@ndefined
                                    \hyperstr@pblank"#2 \hyper@nd"\else
                                    #1\hyperstr@pblank" #2 \hyper@nd"\fi\fi}
\long\def\hyper@nchor#1#2{\edef\hyperm@cro{html:<A #1>}%
                          \special\expandafter{\hyperm@cro}%
                          {#2}}%
\def\hyper@atm@ning#1->#2\hyper@nd{#2}
\def\hyperlink#1{\edef\hypert@mp{#1}%
               \edef\hypert@mp{\expandafter\hyper@atm@ning\meaning\hypert@mp
                               \hyper@nd}%
               \edef\hypert@mp"{ \expandafter\hyperstr@pquote\expandafter"%
                               \hypert@mp"\hyper@nd}%
               \edef\hypert@mp{\expandafter\hyperstr@pblank\expandafter%
                               "\hypert@mp" \hyper@nd"}%
               \hyper@nchor{href=\expandafter"\hypert@mp"}}%
\def\hypertarget#1{\edef\hypert@mp{#1}%
               \edef\hypert@mp{\expandafter\hyper@atm@ning\meaning\hypert@mp
                               \hyper@nd}%
               \edef\hypert@mp"{ \expandafter\hyperstr@pquote\expandafter"%
                               \hypert@mp"\hyper@nd}%
               \edef\hypert@mp{\expandafter\hyperstr@pblank\expandafter%
                               "\hypert@mp" \hyper@nd"}%
               \hyper@nchor{name=\expandafter"\hypert@mp"}}%
\def\hyperref{\afterassignment\hyperr@f\let\hyperp@ram}
\def\hyperr@f{\ifx\hyperp@ram{\iffalse}\fi
               \expandafter\expandafter\expandafter\hyperr@@
               \expandafter{%
              \else
               \iffalse}\fi
               \ifx\hyperp@ram\hyper@ndefined
                 \message{Undefined reference}%
                 \def\hyperp@r@m{{}{undefined}{}}%
               \else
                 \edef\hyperp@r@m{\hyperp@ram}%
               \fi
               \expandafter\expandafter\expandafter\hyperr@@
               \expandafter\hyperp@r@m
              \fi}%
% refer to #1, \hyperh@sh#2.#3 or #1\hyperh@sh#2.#3
% depending on what is blank/nonblank
\def\hyperr@@#1#2#3{\ifx\hyper@ndefined#1\hyper@ndefined
                    \hypert@ks\expandafter{\hyperh@sh#2.#3}%
                    \else
                     \ifx\hyper@ndefined#2#3\hyper@ndefined
                      \hypert@ks{#1}%
                     \else
                      \def\hypert@mp{#1}%
                      \hypert@ks\expandafter\expandafter\expandafter
                      {\expandafter\hypert@mp\hyperh@sh#2.#3}%
                     \fi
                    \fi
                    \expandafter\hyperlink\expandafter{\the\hypert@ks}}%
\def\hyperdef#1#2#3{{\global\escapechar=`\\\relax
                     \edef\hypert@mp{\hyperstr@pquote"#2.#3"\hyper@nd}%
                     \expandafter\ifx\csname hyperd@\meaning\hypert@mp
                     \endcsname
                     \relax
                     \expandafter\gdef\csname hyperd@\meaning\hypert@mp
                     \endcsname{}%
                     \gdef#1{{}{\hyperstr@pquote"#2"\hyper@nd}%
                               {\hyperstr@pquote"#3"\hyper@nd}}%
                     \immediate\write\hyperf@le{\def\noexpand#1{#1}}%
                     \xdef\hypert@mp{\global\let\noexpand\hypert@mp=\relax
                                     \noexpand\hypertarget{\hypert@mp}}%
                     \global\hypert@ks={\hypert@mp}%
                     \else
                     \message\expandafter{'\hypert@mp' duplicate}%
                     \global\let\hypert@mp=\relax
                     \global\hypert@ks={\hyperdef{#1}{#2}{#3@}}%
                     \fi}\the\hypert@ks}%

\def\hyper@nique#1#2#3#4{\global\escapechar=`\\\relax
                     \edef\hypert@mp{\hyperstr@pquote"#2.#3"\hyper@nd}%
                     \expandafter\ifx\csname hyperd@\meaning\hypert@mp
                     \endcsname
                     \relax
                     \gdef#1{{}{\hyperstr@pquote"#2"\hyper@nd}%
                               {\hyperstr@pquote"#3"\hyper@nd}}%
                     \global\let\hypert@mp=\relax
                     #4%
                     \else
                     \global\let\hypert@mp=\relax
                     \hyper@nique{#1}{#2}{#3@}{#4}%
                     \fi
                     }%

%%% 
% protection macros
%%%
\let\hyper@@@@=\relax
\def\hyper@@{\let\hyper@@@=\relax}%
\hyper@@
\def\hyper@{\relax\let\hyper@@@\noexpand\hyper@\noexpand}%
\def\hyperpr@ref{\hyper@@\hyperref}
\def\hyperpr@def{\hyper@@\hyperdef}

% As per pg's suggestion
\let\href\hyperlink

%
% Restore the catcode of @
%
\hypers@fe

\catcode`\@=11
%%% saclay A4 paper:
\def\unredoffs{\voffset=13mm \hoffset=6.5truemm} 
\def\redoffs{\voffset=-12.truemm\hoffset=-9truemm} 
\def\speclscape{}
%
%---------------------------------------------------------------------%
\newbox\leftpage \newdimen\fullhsize \newdimen\hstitle \newdimen\hsbody
\newdimen\hdim
%\tolerance=1000
\hfuzz=1pt
\ifx\hyperdef\UNd@FiNeD\def\hyperdef#1#2#3#4{#4}\def\hyperref#1#2#3#4{#4}\fi
\def\newans{y }
%\message{ new or old (y/n)? }\read-1 to\answb
\def\answb{y }
\ifx\answb\newans\message{(This uses normal fonts.)}%
%\magnification=1200
%\hoffset=6.7mm
%\voffset=13mm
% ***************************************************************************
%
\def\bigans{b }
%\message{ big or little (b/l)? }\read-1 to\answ
%
\def\answ{b }
\ifx\answ\bigans\message{(Format simple colonne 12pts.}
\magnification=1200 \unredoffs\hsize=147truemm\vsize=219truemm
\hsbody=\hsize \hstitle=\hsize %take default values for unreduced format
\else\message{(Format double colonne, 10pts.} \let\l@r=L
\magnification=1000 \vsize=182.5truemm
\redoffs%
\hstitle=122.5truemm\hsbody=122.5truemm\fullhsize=261.5truemm\hsize=\hsbody 
\output={
  \almostshipout{\leftline{\vbox{\makeheadline\pagebody\makefootline}}}
\advancepageno%
%  \fi
}
\def\almostshipout#1{\if L\l@r \count1=1 \message{[\the\count0.\the\count1]}
      \global\setbox\leftpage=#1 \global\let\l@r=R
 \else \count1=2
  \shipout\vbox{\speclscape{\hsize\fullhsize}%\makeheadline}
      \hbox to\fullhsize{\box\leftpage\hfil#1}}  \global\let\l@r=L\fi}
\fi
% ****************************************************************************
% fonts, Dirac slash

\def\sla#1{\mkern-1.5mu\raise0.4pt\hbox{$\not$}\mkern1.2mu #1\mkern 0.7mu}
\def\Dbar{\mkern-1.5mu\raise0.4pt\hbox{$\not$}\mkern-.1mu {\rm D}\mkern.1mu}
\def\Abar{\mkern1.mu\raise0.4pt\hbox{$\not$}\mkern-1.3mu A\mkern.1mu}
\def\Bbar{\mkern-0.mu\raise0.4pt\hbox{$\not$}\mkern-.3mu B\mkern 0.6mu}
\newskip\tableskipamount \tableskipamount=8pt plus 3pt minus 3pt
\def\tableskip{\vskip\tableskipamount}
%****************************

\newdimen\chapskip

\font\ssbx=cmssbx10  

\font\caprm=cmr9
\font\capit=cmti9
\font\capbf=cmbx9
\font\capsl=cmsl9
\font\capmi=cmmi9
\font\capex=cmex9
\font\capsy=cmsy9
\chapskip=17.5mm
\def\makeheadline{\vbox to 0pt{\vskip-22.5pt
\line{\vbox to8.5pt{}\the\headline}\vss}\nointerlineskip}
%***************************************************
  %obsolete??
\font\tenbi=cmmib10 
\font\ninebi=cmmib9
\font\sevenbi=cmmib7 
\font\fivebi=cmmib5
\textfont4=\tenbi
\scriptfont4=\sevenbi
\scriptscriptfont4=\fivebi
\font\headrm=cmr10

%****************************
\font\sixrm=cmr6
\font\fiverm=cmr5
\font\sixmi=cmmi6
\font\fivemi=cmmi5
\font\sixsy=cmsy6
\font\fivesy=cmsy5
\font\sixbf=cmbx6
\font\fivebf=cmbx5
\skewchar\capmi='177 \skewchar\sixmi='177 \skewchar\fivemi='177
\skewchar\capsy='60 \skewchar\sixsy='60 \skewchar\fivesy='60

\def\elevenpoint{
\textfont0=\caprm \scriptfont0=\sixrm \scriptscriptfont0=\fiverm
\def\rm{\fam0\caprm}
\textfont1=\capmi \scriptfont1=\sixmi \scriptscriptfont1=\fivemi
\textfont2=\capsy \scriptfont2=\sixsy \scriptscriptfont2=\fivesy
\textfont3=\capex \scriptfont3=\capex \scriptscriptfont3=\capex
\textfont\itfam=\capit \def\it{\fam\itfam\capit} % \it is family 4
\textfont\slfam=\capsl  \def\sl{\fam\slfam\capsl} % \sl is family 5
\textfont\bffam=\capbf \scriptfont\bffam=\sixbf
\scriptscriptfont\bffam=\fivebf
\def\bf{\fam\bffam\capbf} % \bf is family 6
\textfont4=\ninebi \scriptfont4=\sevenbi \scriptscriptfont4=\fivebi
\abovedisplayskip=11pt plus 3pt minus 8pt
\belowdisplayskip=\abovedisplayskip
\smallskipamount=2.7pt plus 1pt minus 1pt
\medskipamount=5.4pt plus 2pt minus 2pt
\bigskipamount=10.8pt plus 3.6pt minus 3.6pt
\normalbaselineskip=11pt
\setbox\strutbox=\hbox{\vrule height7.8pt depth3.2pt width0pt}
\normalbaselines \rm}
% *************************************************************************

% ****************************************************************************
%		*****	  MSSYMB.TeX	*****		       20 Sept 91
%
%	This file contains the definitions for the symbols in the two
%	"extra symbols" fonts created at the American Math. Society.
%
%       The old fonts msxm et msym have been replaced by msam et msbm. 

\catcode`\@=11

\font\tenmsa=msam10
\font\sevenmsa=msam7
\font\fivemsa=msam5
\font\tenmsb=msbm10
\font\sevenmsb=msbm7
\font\fivemsb=msbm5
\newfam\msafam
\newfam\msbfam
\textfont\msafam=\tenmsa  \scriptfont\msafam=\sevenmsa
  \scriptscriptfont\msafam=\fivemsa
\textfont\msbfam=\tenmsb  \scriptfont\msbfam=\sevenmsb
  \scriptscriptfont\msbfam=\fivemsb

\def\hexnumber@#1{\ifcase#1 0\or1\or2\or3\or4\or5\or6\or7\or8\or9\or
	A\or B\or C\or D\or E\or F\fi }

%  The following 13 lines establish the use of the Euler Fraktur font.
%  To use this font, remove % from beginning of these lines.
\font\teneuf=eufm10
\font\seveneuf=eufm7
\font\fiveeuf=eufm5
\newfam\euffam
\textfont\euffam=\teneuf
\scriptfont\euffam=\seveneuf
\scriptscriptfont\euffam=\fiveeuf
\def\frak{\ifmmode\let\next\frak@\else
 \def\next{\Err@{Use \string\frak\space only in math mode}}\fi\next}
\def\goth{\ifmmode\let\next\frak@\else
 \def\next{\Err@{Use \string\goth\space only in math mode}}\fi\next}
\def\frak@#1{{\frak@@{#1}}}
\def\frak@@#1{\fam\euffam#1}
%  End definition of Euler Fraktur font.

\edef\msa@{\hexnumber@\msafam}
\edef\msb@{\hexnumber@\msbfam}

\mathchardef\boxdot="2\msa@00
\mathchardef\boxplus="2\msa@01
\mathchardef\boxtimes="2\msa@02
\mathchardef\square="0\msa@03
\mathchardef\blacksquare="0\msa@04
\mathchardef\centerdot="2\msa@05
\mathchardef\lozenge="0\msa@06
\mathchardef\blacklozenge="0\msa@07
\mathchardef\circlearrowright="3\msa@08
\mathchardef\circlearrowleft="3\msa@09
\mathchardef\rightleftharpoons="3\msa@0A
\mathchardef\leftrightharpoons="3\msa@0B
\mathchardef\boxminus="2\msa@0C
\mathchardef\Vdash="3\msa@0D
\mathchardef\Vvdash="3\msa@0E
\mathchardef\vDash="3\msa@0F
\mathchardef\twoheadrightarrow="3\msa@10
\mathchardef\twoheadleftarrow="3\msa@11
\mathchardef\leftleftarrows="3\msa@12
\mathchardef\rightrightarrows="3\msa@13
\mathchardef\upuparrows="3\msa@14
\mathchardef\downdownarrows="3\msa@15
\mathchardef\upharpoonright="3\msa@16

\mathchardef\downharpoonright="3\msa@17
\mathchardef\upharpoonleft="3\msa@18
\mathchardef\downharpoonleft="3\msa@19
\mathchardef\rightarrowtail="3\msa@1A
\mathchardef\leftarrowtail="3\msa@1B
\mathchardef\leftrightarrows="3\msa@1C
\mathchardef\rightleftarrows="3\msa@1D
\mathchardef\Lsh="3\msa@1E
\mathchardef\Rsh="3\msa@1F
\mathchardef\rightsquigarrow="3\msa@20
\mathchardef\leftrightsquigarrow="3\msa@21
\mathchardef\looparrowleft="3\msa@22
\mathchardef\looparrowright="3\msa@23
\mathchardef\circeq="3\msa@24
\mathchardef\succsim="3\msa@25
\mathchardef\gtrsim="3\msa@26
\mathchardef\gtrapprox="3\msa@27
\mathchardef\multimap="3\msa@28
\mathchardef\therefore="3\msa@29
\mathchardef\because="3\msa@2A
\mathchardef\doteqdot="3\msa@2B

\mathchardef\triangleq="3\msa@2C
\mathchardef\precsim="3\msa@2D
\mathchardef\lesssim="3\msa@2E
\mathchardef\lessapprox="3\msa@2F
\mathchardef\eqslantless="3\msa@30
\mathchardef\eqslantgtr="3\msa@31
\mathchardef\curlyeqprec="3\msa@32
\mathchardef\curlyeqsucc="3\msa@33
\mathchardef\preccurlyeq="3\msa@34
\mathchardef\leqq="3\msa@35
\mathchardef\leqslant="3\msa@36
\mathchardef\lessgtr="3\msa@37
\mathchardef\backprime="0\msa@38
\mathchardef\risingdotseq="3\msa@3A
\mathchardef\fallingdotseq="3\msa@3B
\mathchardef\succcurlyeq="3\msa@3C
\mathchardef\geqq="3\msa@3D
\mathchardef\geqslant="3\msa@3E
\mathchardef\gtrless="3\msa@3F
\mathchardef\sqsubset="3\msa@40
\mathchardef\sqsupset="3\msa@41
\mathchardef\vartriangleright="3\msa@42
\mathchardef\vartriangleleft="3\msa@43
\mathchardef\trianglerighteq="3\msa@44
\mathchardef\trianglelefteq="3\msa@45
\mathchardef\bigstar="0\msa@46
\mathchardef\between="3\msa@47
\mathchardef\blacktriangledown="0\msa@48
\mathchardef\blacktriangleright="3\msa@49
\mathchardef\blacktriangleleft="3\msa@4A
\mathchardef\vartriangle="0\msa@4D
\mathchardef\blacktriangle="0\msa@4E
\mathchardef\triangledown="0\msa@4F
\mathchardef\eqcirc="3\msa@50
\mathchardef\lesseqgtr="3\msa@51
\mathchardef\gtreqless="3\msa@52
\mathchardef\lesseqqgtr="3\msa@53
\mathchardef\gtreqqless="3\msa@54
\mathchardef\Rrightarrow="3\msa@56
\mathchardef\Lleftarrow="3\msa@57
\mathchardef\veebar="2\msa@59
\mathchardef\barwedge="2\msa@5A
\mathchardef\doublebarwedge="2\msa@5B
\mathchardef\angle="0\msa@5C
\mathchardef\measuredangle="0\msa@5D
\mathchardef\sphericalangle="0\msa@5E
\mathchardef\varpropto="3\msa@5F
\mathchardef\smallsmile="3\msa@60
\mathchardef\smallfrown="3\msa@61
\mathchardef\Subset="3\msa@62
\mathchardef\Supset="3\msa@63
\mathchardef\Cup="2\msa@64

\mathchardef\Cap="2\msa@65

\mathchardef\curlywedge="2\msa@66
\mathchardef\curlyvee="2\msa@67
\mathchardef\leftthreetimes="2\msa@68
\mathchardef\rightthreetimes="2\msa@69
\mathchardef\subseteqq="3\msa@6A
\mathchardef\supseteqq="3\msa@6B
\mathchardef\bumpeq="3\msa@6C
\mathchardef\Bumpeq="3\msa@6D
\mathchardef\lll="3\msa@6E

\mathchardef\ggg="3\msa@6F

\mathchardef\circledS="0\msa@73
\mathchardef\pitchfork="3\msa@74
\mathchardef\dotplus="2\msa@75
\mathchardef\backsim="3\msa@76
\mathchardef\backsimeq="3\msa@77
\mathchardef\complement="0\msa@7B
\mathchardef\intercal="2\msa@7C
\mathchardef\circledcirc="2\msa@7D
\mathchardef\circledast="2\msa@7E
\mathchardef\circleddash="2\msa@7F
\def\ulcorner{\delimiter"4\msa@70\msa@70 }
\def\urcorner{\delimiter"5\msa@71\msa@71 }
\def\llcorner{\delimiter"4\msa@78\msa@78 }
\def\lrcorner{\delimiter"5\msa@79\msa@79 }
\def\yen{\mathhexbox\msa@55 }
\def\checkmark{\mathhexbox\msa@58 }
\def\circledR{\mathhexbox\msa@72 }
\def\maltese{\mathhexbox\msa@7A }
\mathchardef\lvertneqq="3\msb@00
\mathchardef\gvertneqq="3\msb@01
\mathchardef\nleq="3\msb@02
\mathchardef\ngeq="3\msb@03
\mathchardef\nless="3\msb@04
\mathchardef\ngtr="3\msb@05
\mathchardef\nprec="3\msb@06
\mathchardef\nsucc="3\msb@07
\mathchardef\lneqq="3\msb@08
\mathchardef\gneqq="3\msb@09
\mathchardef\nleqslant="3\msb@0A
\mathchardef\ngeqslant="3\msb@0B
\mathchardef\lneq="3\msb@0C
\mathchardef\gneq="3\msb@0D
\mathchardef\npreceq="3\msb@0E
\mathchardef\nsucceq="3\msb@0F
\mathchardef\precnsim="3\msb@10
\mathchardef\succnsim="3\msb@11
\mathchardef\lnsim="3\msb@12
\mathchardef\gnsim="3\msb@13
\mathchardef\nleqq="3\msb@14
\mathchardef\ngeqq="3\msb@15
\mathchardef\precneqq="3\msb@16
\mathchardef\succneqq="3\msb@17
\mathchardef\precnapprox="3\msb@18
\mathchardef\succnapprox="3\msb@19
\mathchardef\lnapprox="3\msb@1A
\mathchardef\gnapprox="3\msb@1B
\mathchardef\nsim="3\msb@1C
%\mathchardef\napprox="3\msb@1D
\mathchardef\ncong="3\msb@1D

\mathchardef\varsubsetneq="3\msb@20
\mathchardef\varsupsetneq="3\msb@21
\mathchardef\nsubseteqq="3\msb@22
\mathchardef\nsupseteqq="3\msb@23
\mathchardef\subsetneqq="3\msb@24
\mathchardef\supsetneqq="3\msb@25
\mathchardef\varsubsetneqq="3\msb@26
\mathchardef\varsupsetneqq="3\msb@27
\mathchardef\subsetneq="3\msb@28
\mathchardef\supsetneq="3\msb@29
\mathchardef\nsubseteq="3\msb@2A
\mathchardef\nsupseteq="3\msb@2B
\mathchardef\nparallel="3\msb@2C
\mathchardef\nmid="3\msb@2D
\mathchardef\nshortmid="3\msb@2E
\mathchardef\nshortparallel="3\msb@2F
\mathchardef\nvdash="3\msb@30
\mathchardef\nVdash="3\msb@31
\mathchardef\nvDash="3\msb@32
\mathchardef\nVDash="3\msb@33
\mathchardef\ntrianglerighteq="3\msb@34
\mathchardef\ntrianglelefteq="3\msb@35
\mathchardef\ntriangleleft="3\msb@36
\mathchardef\ntriangleright="3\msb@37
\mathchardef\nleftarrow="3\msb@38
\mathchardef\nrightarrow="3\msb@39
\mathchardef\nLeftarrow="3\msb@3A
\mathchardef\nRightarrow="3\msb@3B
\mathchardef\nLeftrightarrow="3\msb@3C
\mathchardef\nleftrightarrow="3\msb@3D
\mathchardef\divideontimes="2\msb@3E
\mathchardef\varnothing="0\msb@3F
\mathchardef\nexists="0\msb@40
\mathchardef\mho="0\msb@66
\mathchardef\eth="0\msb@67
\mathchardef\eqsim="3\msb@68
\mathchardef\beth="0\msb@69
\mathchardef\gimel="0\msb@6A
\mathchardef\daleth="0\msb@6B
\mathchardef\lessdot="3\msb@6C
\mathchardef\gtrdot="3\msb@6D
\mathchardef\ltimes="2\msb@6E
\mathchardef\rtimes="2\msb@6F
\mathchardef\shortmid="3\msb@70
\mathchardef\shortparallel="3\msb@71
\mathchardef\smallsetminus="2\msb@72
\mathchardef\thicksim="3\msb@73
\mathchardef\thickapprox="3\msb@74
\mathchardef\approxeq="3\msb@75
\mathchardef\succapprox="3\msb@76
\mathchardef\precapprox="3\msb@77
\mathchardef\curvearrowleft="3\msb@78
\mathchardef\curvearrowright="3\msb@79
\mathchardef\digamma="0\msb@7A
\mathchardef\varkappa="0\msb@7B
\mathchardef\hslash="0\msb@7D
\mathchardef\hbar="0\msb@7E
\mathchardef\backepsilon="3\msb@7F
\def\Bbb{\ifmmode\let\next\Bbb@\else
 \def\next{\errmessage{Use \string\Bbb\space only in math mode}}\fi\next}
\def\Bbb@#1{{\Bbb@@{#1}}}
\def\Bbb@@#1{\fam\msbfam#1}
\font\sacfont=eufm10 scaled 1440
\catcode`\@=12

\def\sla#1{\mkern-1.5mu\raise0.4pt\hbox{$\not$}\mkern1.2mu #1\mkern 0.7mu}
\def\Dbar{\mkern-1.5mu\raise0.4pt\hbox{$\not$}\mkern-.1mu {\rm D}\mkern.1mu}
\def\Abar{\mkern1.mu\raise0.4pt\hbox{$\not$}\mkern-1.3mu A\mkern.1mu}
\nopagenumbers
\headline={\ifnum\pageno=1\hfill\else\draftdate\hfil{\headrm\folio}%
\hfil\fi}	 
\else\message{(This uses pseudo 12pts fonts.}
\hoffset=8mm
\voffset=16mm
\input lfont12 %pour sun

\def\sla#1{\mkern-1.5mu\raise0.5pt\hbox{$\not$}\mkern1.2mu #1\mkern 0.7mu}
\def\Dbar{\mkern-1.5mu\raise0.5pt\hbox{$\not$}\mkern-.1mu {\rm D}\mkern.1mu}
\def\Abar{\mkern1.mu\raise0.5pt\hbox{$\not$}\mkern-1.3mu A\mkern.1mu}
\fi
% ****************************************************************************
%********* end ouptut macros
% ****************************************************************************
% ****** extrait de definit.tex (obsolete ?)
\def\fileth{\noalign{\hrule}}

% ****************************************************************************
\newcount\yearltd\yearltd=\year\advance\yearltd by -1900
\newif\ifdraftmode
\draftmodefalse
\def\draft{\draftmodetrue{\count255=\time\divide\count255 by 60
\xdef\hourmin{\number\count255} 
  \multiply\count255 by-60\advance\count255 by\time
  \xdef\hourmin{\hourmin:\ifnum\count255<10 0\fi\the\count255}}}
\def\draftdate{\ifdraftmode{\headrm\quad (le
\number\day/\number\month/\number\yearltd\ \ \hourmin)}\else{}\fi} 
\newif\iffrancmode
\francmodefalse
% ********* A few math symbols
\def\e{\mathop{\rm e}\nolimits}

\def\d{{\rm d}}
\def\ud{{\textstyle{1\over 2}}}

\def\del{\partial}

\chardef\sigmat=27

\def\frac#1#2{{\textstyle{#1\over#2}}}

\def\today{\number\day/\number\month/\number\year}
\def\leaderfill{\leaders\hbox to 1em{\hss.\hss}\hfill}
% *************************************************************************
\catcode`\@=11
% ************** double alignment in eqalignno style **********************
\def\deqalignno#1{\displ@y\tabskip\centering \halign to
\displaywidth{\hfil$\displaystyle{##}$\tabskip0pt&$\displaystyle{{}##}$
\hfil\tabskip0pt &\quad
\hfil$\displaystyle{##}$\tabskip0pt&$\displaystyle{{}##}$ 
\hfil\tabskip\centering& \llap{$##$}\tabskip0pt \crcr #1 \crcr}}
% ************** double eqalign ******************************************
\def\deqalign#1{\null\,\vcenter{\openup\jot\m@th\ialign{
\strut\hfil$\displaystyle{##}$&$\displaystyle{{}##}$\hfil
&&\quad\strut\hfil$\displaystyle{##}$&$\displaystyle{{}##}$
\hfil\crcr#1\crcr}}\,}
%***************************************************************************
%********* titlepage, headline, section, subsection, sub, appendix *********
%***************************************************************************
%**************** input with check of file existence ***********************
\newread\ch@ckfile
\def\cinput#1{\def\filen@me{#1}%                              DOES NOT WORK???
\immediate\openin\ch@ckfile=\filen@me
\ifeof\ch@ckfile\closein\ch@ckfile\message{<< (\filen@me\ N'EXISTE PAS)
>>}\else% 
\input\filen@me\closein \ch@ckfile\fi}
%********* introduce equation number file: for non-causal quotation
\immediate\openin\ch@ckfile=\jobname.def
\ifeof\ch@ckfile\closein\ch@ckfile\message{<< (\jobname.def N'EXISTE PAS)
>>}
% Warning macro
\def\DefWarn#1{\ifx\UNd@FiNeD#1\else
\immediate\write16{*** WARNING: the label \string#1 is already defined%
***}\fi}% 
\else% 
\def\DefWarn#1{}
\input\jobname.def\closein \ch@ckfile\fi
%**************************************************************************
\newcount\nosection
\newcount\nosubsection
\newcount\neqno
\newcount\notenumber
\newcount\nofigure
\newif\ifappmode
\def\table{\jobname.toc}
\def\equation{\jobname.equ}
\def\labeldefs{\jobname.eqd}
\newwrite\equa
\newwrite\tab 
\newwrite\eqdf
% ******************* titlepage **********************************

\newdimen\hulp
\def\maketitle#1{
\edef\oneliner##1{\centerline{##1}}
\edef\twoliner##1{\vbox{\parindent=0pt\leftskip=0pt plus 1fill\rightskip=0pt
plus 1fill 
                     \parfillskip=0pt\relax##1}} 
\setbox0=\vbox{#1}\hulp=0.5\hsize
                 \ifdim\wd0<\hulp\oneliner{#1}\else
                 \twoliner{#1}\fi}
\def\preprint#1{{\sacfont }\hfill{#1}\vskip 20mm}
% **************** beginning
\def\title#1\par{\gdef\titlename{#1}
\maketitle{\ssbx\uppercase\expandafter{\titlename}}
\vskip20truemm
\nosection=0
\neqno=0
\notenumber=0
\nofigure=0
\def\prefix{}
\appmodefalse
\mark{\the\nosection}
\message{#1}
%\immediate\openout\tab=\table
\immediate\openout\equa=\equation
\immediate\openout\eqdf=\labeldefs
}
\def\abstract{\vskip8mm\iffrancmode\centerline{RESUME}\else%
\centerline{ABSTRACT}\fi \smallskip \begingroup\narrower
\elevenpoint\baselineskip10pt} 
\def\endabstract{\par\endgroup \bigskip}
% ***************** input table of contents
%\def\content{\vfill\eject% A un bug dans le format double colonne
%\begingroup\centerline{\uppercase\expandafter{Table of contents}}% 
%\bigskip\elevenpoint\noindent% 
%\parindent=2em
%\openin 1=\jobname.tab
%\ifeof 1\closein1\message{<< (\jobname.tab DOES NOT EXIST. TeX again)>>}%
%\else\input\jobname.tab\closein 1\fi 
%\endgroup\vfill\eject}
%******************************* section ***********************************
\def\section#1\par{\vskip0pt plus.1\vsize\penalty-100\vskip0pt plus-.1
\vsize\bigskip\vskip\parskip
\ifnum\nosection=0\ifappmode\relax\else\writetoc
\fi\fi% ajout
\advance\nosection by 1\global\nosubsection=0\global\neqno=0
\vbox{\noindent\bf{\hyperdef\hypernoname{section}{\prefix\the\nosection}%
{\prefix\the\nosection}\ #1}}
\writetoca{{\string\hyperref{}{section}{\prefix\the\nosection}%
{\prefix\the\nosection}} {#1}}
\message{\the\nosection\ #1}
\mark{\the\nosection}\bigskip\noindent
%\xdef\ecrire{\write\tab{\string\par\string\item{\prefix\the\nosection}
%#1\string\leaderfill {\noexpand\number\pageno}}}\ecrire
}

% appendix
\def\appendix#1#2\par{\bigbreak\nosection=0
\appmodetrue
\notenumber=0
\neqno=0
\def\prefix{A}
\mark{\the\nosection}
\message{APPENDICES}
{\leftline{APPENDICES} \hyperdef\hypernoname{appendix}{app}{ 
\leftline{\uppercase\expandafter{#1}}
\leftline{\uppercase\expandafter{#2}}}}
\bigskip\noindent\nonfrenchspacing
\writetoca{\string\hyperref{}{appendix}{app}{Appendices}.\ #1.\ #2}%
}
% **************************** \subsection *************************
\def\subsection#1\par {\vskip0pt plus.05\vsize\penalty-100\vskip0pt
plus-.05\vsize\bigskip\vskip\parskip\advance\nosubsection by 1
\vbox{\noindent\it{\hyperdef\hypernoname{subsection}{\prefix\the\nosection.%
\the\nosubsection}{\prefix\the\nosection.\the\nosubsection\ #1}}}%
\smallskip\noindent 
\writetoca{{\string\hyperref{}{subsection}{\prefix\the\nosection.%
\the\nosubsection}{\prefix\the\nosection.\the\nosubsection}} {#1}}
\message{\the\nosection.\the\nosubsection\ #1}
} 
\def\note #1{\advance\notenumber by 1
\footnote{$^{\the\notenumber}$}{\sevenrm #1}} 
% ?????

%\def\enchapter{\end}
\parindent=1em 
\newinsert\margin
\dimen\margin=\maxdimen
\count\margin=0 \skip\margin=0pt
%*****************************************************************
% IMPORTANT, new version demands chapter be defined before any section,
% section be defined before any subsection
\def\sslbl#1{\DefWarn#1%
\ifdraftmode{\hfill\escapechar-1{\rlap{\hskip-1mm%
\sevenrm\string#1}}}\fi%
\ifnum\nosection=0\xdef#1{}%
\edef\ewrite{\write\eqdf{\string\def\string#1{}}
\write\eqdf{}}\ewrite%
\edef\ewrite{\write\equa{{\string#1}}%
\write\equa{}}\ewrite%
\else%
\ifnum\nosubsection=0%
%\xdef#1{\prefix\the\nosection}%
\xdef#1{\noexpand\hyperref{}{section}{\prefix\the\nosection}{\prefix\the\nosection}}%
\edef\ewrite{\write\eqdf{\string\def\string#1{\prefix%
\the\nosection}}\write\eqdf{}}\ewrite%
\edef\ewrite{\write\equa{{\string#1},\prefix\the\nosection}%
\write\equa{}}\ewrite%
\writedef{#1\leftbracket#1}
\else%
%\xdef#1{\prefix\the\nosection.\the\nosubsection}%
\xdef#1{\noexpand\hyperref{}{subsection}{\prefix\the\nosection.%
\the\nosubsection}{\prefix\the\nosection.\the\nosubsection}}%
\writedef{#1\leftbracket#1}
\edef\ewrite{\write\eqdf{\string\def\string#1{\prefix%
\the\nosection.\the\nosubsection}}\write\eqdf{}}\ewrite%
\edef\ewrite{\write\equa{{\string#1},\prefix\the\nosection%
.\the\nosubsection}\write\equa{}}\ewrite\fi\fi}%
%**********************************************************************
% Autre utilitaire
\def\listcontent{\immediate\closeout\tfile\immediate\openin%
\ch@ckfile=\jobname.tab
\ifeof\ch@ckfile\message{no file \jobname.tab, no table of contents this
pass}%
\else\closein\ch@ckfile\centerline{\bf\iffrancmode Table des
mati\`eres \else Contents\fi}\nobreak\medskip% 
{\baselineskip=12pt\parskip=0pt\catcode`\@=11\input\jobname.tab
\catcode`\@=12\bigbreak\bigskip}\fi}
\newwrite\tfile \def\writetoca#1{}
%       use this to write file with table of contents
\def\writetoc{\immediate\openout\tfile=\jobname.tab
   \def\writetoca##1{{\edef\next{\write\tfile{\noindent ##1
   \string\leaderfill {\string\hyperref{}{page}{\noexpand\number\pageno}%
                       {\noexpand\number\pageno}} \par}}\next}}}

% ********************* references harvmac style ***********************
%     \ref\label{text}
% generates a number, assigns it to \label, generates an entry.
% To list the refs on a separate page,  \listrefs
%
\def\nolabels{\def\wrlabeL##1{}\def\eqlabeL##1{}\def\reflabeL##1{}}
\def\writelabels{\def\wrlabeL##1{\leavevmode\vadjust{\rlap{\smash%
{\line{{\escapechar=` \hfill\rlap{\sevenrm\hskip.03in\string##1}}}}}}}%
\def\eqlabeL##1{{\escapechar-1\rlap{\sevenrm\hskip.05in\string##1}}}%
\def\reflabeL##1{\noexpand\llap{\noexpand\sevenrm\string\string\string##1}}}
\nolabels

\global\newcount\refno \global\refno=1
\newwrite\rfile
\def\ref{[\hyperref{}{reference}{\the\refno}{\the\refno}]\nref}
\def\nref#1{\DefWarn#1%
\xdef#1{[\noexpand\hyperref{}{reference}{\the\refno}{\the\refno}]}%
\writedef{#1\leftbracket#1}%
\ifnum\refno=1\immediate\openout\rfile=\jobname.ref\fi
\chardef\wfile=\rfile\immediate\write\rfile{\noexpand\item{[\noexpand\hyperdef%
\noexpand\hypernoname{reference}{\the\refno}{\the\refno}]\ }%
\reflabeL{#1\hskip.31in}\pctsign}\global\advance\refno by1\findarg}
%       horrible hack to sidestep tex \write limitation
\def\findarg#1#{\begingroup\obeylines\newlinechar=`\^^M\pass@rg}
{\obeylines\gdef\pass@rg#1{\writ@line\relax #1^^M\hbox{}^^M}%
\gdef\writ@line#1^^M{\expandafter\toks0\expandafter{\striprel@x #1}%
\edef\next{\the\toks0}\ifx\next\em@rk\let\next=\endgroup\else\ifx\next\empty%
\else\immediate\write\wfile{\the\toks0}\fi\let\next=\writ@line\fi\next\relax}}
\def\striprel@x#1{} \def\em@rk{\hbox{}}
\def\lref{\begingroup\obeylines\lr@f}
\def\lr@f#1#2{\DefWarn#1\gdef#1{\let#1=\UNd@FiNeD\ref#1{#2}}\endgroup\unskip}
\def\semi{;\hfil\break}
\def\addref#1{\immediate\write\rfile{\noexpand\item{}#1}} %now unnecessary
\def\listrefs{{}\vfill\supereject\immediate\closeout\rfile\writestoppt
\baselineskip=14pt\centerline{{\bf\iffrancmode R\'eferences\else References%
\fi}}
\bigskip{\parindent=20pt%
\frenchspacing\escapechar=` \input \jobname.ref\vfill\eject}\nonfrenchspacing}
\def\startrefs#1{\immediate\openout\rfile=\jobname.ref\refno=#1}
\def\xref{\expandafter\xr@f}\def\xr@f[#1]{#1}
\def\refs#1{\count255=1[\r@fs #1{\hbox{}}]}
\def\r@fs#1{\ifx\UNd@FiNeD#1\message{reflabel \string#1 is undefined.}%
\nref#1{need to supply reference \string#1.}\fi%
\vphantom{\hphantom{#1}}{\let\hyperref=\relax\xdef\next{#1}}%
\ifx\next\em@rk\def\next{}%
\else\ifx\next#1\ifodd\count255\relax\xref#1\count255=0\fi%
\else#1\count255=1\fi\let\next=\r@fs\fi\next}
%************************
%*******
%
\newwrite\lfile
{\escapechar-1\xdef\pctsign{\string\%}\xdef\leftbracket{\string\{}
\xdef\rightbracket{\string\}}\xdef\numbersign{\string\#}}
\def\writedefs{\immediate\openout\lfile=\jobname.def \def\writedef##1{%
{\let\hyperref=\relax\let\hyperdef=\relax\let\hypernoname=\relax
 \immediate\write\lfile{\string\def\string##1\rightbracket}}}}%
\def\writestop{\def\writestoppt{\immediate\write\lfile{\string\pageno%
\the\pageno\string\startrefs\leftbracket\the\refno\rightbracket%
\string\def\string\secsym\leftbracket\secsym\rightbracket%
\string\secno\the\secno\string\meqno\the\meqno}\immediate\closeout\lfile}}
\def\writestoppt{}\def\writedef#1{}
\writedefs
% ******
% bibliography: not very satisfactory
\def\biblio\par{\vskip0pt plus.1\vsize\penalty-100\vskip0pt plus-.1
\vsize\bigskip\vskip\parskip
\message{Bibliographie}
{\leftline{\bf \hyperdef\hypernoname{biblio}{bib}{Bibliographical Notes}}}
\nobreak\medskip\noindent\frenchspacing
\writetoca{\string\hyperref{}{biblio}{bib}{Bibliographical Notes}}}%

%**************** autre version si plusieurs biblio ************************
\def\biblionote{\iffrancmode Notes Bibliographiques\else Bibliographical Notes
\fi}
\def\beginbib\par{\vskip0pt plus.1\vsize\penalty-100\vskip0pt plus-.1
\vsize\bigskip\vskip\parskip
\message{Bibliographie}
{\leftline{\bf \hyperdef\hypernoname{biblio}{\the\nosection}%
{\biblionote}}}
\nobreak\medskip\noindent\frenchspacing
\writetoca{\string\hyperref{}{biblio}{\the\nosection}%
{\biblionote}}}%

% *************** exercises: same comment
\def\Exercises{\iffrancmode Exercices\else Exercises
\fi}
\def\exerc\par{\vskip0pt plus.1\vsize\penalty-100\vskip0pt plus-.1
\vsize\bigskip\vskip\parskip
\message{Exercises}
{\leftline{\bf \hyperdef\hypernoname{exercise}{\the\nosection}{\Exercises}}}
\nobreak\medskip\noindent\frenchspacing
\writetoca{\string\hyperref{}{exercise}{\the\nosection}{\Exercises}}
}
%*************************************************************************
%Macro de numerotation automatique
%*************************************************************************
% numbering without naming
\def\eqnn{\global\advance\neqno by 1 \ifinner\relax\else%
\eqno\fi(\prefix\the\nosection.\the\neqno)}
%
% numbering and attaching a name: \eqnd{\ename}
\def\eqnd#1{\global\advance\neqno by 1 
{\xdef#1{($\noexpand\hyperref{}{equation}{\prefix\the\nosection.\the\neqno}%
{\prefix\the\nosection.\the\neqno}$)}}%???
\ifinner\relax\else\eqno\fi(\hyperdef\hypernoname{equation}{\prefix\the%
\nosection.\the\neqno}{\prefix\the\nosection.\the\neqno})
\writedef{#1\leftbracket#1}
\ifdraftmode{\escapechar-1{\rlap{\hskip.2mm\sevenrm\string#1}}}\fi
\edef\ewrite{\write\eqdf{\string\def\string#1{($\prefix\the\nosection.%
\the\neqno$)}}%
\write\eqdf{}}\ewrite%
\edef\ewrite{\write\equa{{\string#1},(\prefix\the\nosection.\the\neqno)
{\noexpand\number\pageno}}\write\equa{}}\ewrite}
%
% for eqalignno, allows (1a) (1b)...
\def\checkm@de#1#2{\ifmmode{\def\f@rst##1{##1}\hyperdef\hypernoname{equation}%
{#1}{#2}}\else\hyperref{}{equation}{#1}{#2}\fi}
\def\f@rst#1{\c@t#1a\em@ark}\def\c@t#1#2\em@ark{#1}
\def\eqna#1{\global\advance\neqno by1\ifdraftmode{\hfill%
\escapechar-1{\rlap{\sevenrm\string#1}}}\fi%
\xdef #1##1{(\noexpand\relax\noexpand%
\checkm@de{\prefix\the\nosection.\the\neqno\noexpand\f@rst{##1}1}%
{\hbox{$\prefix\the\nosection.\the\neqno##1$}})}
\writedef{#1\numbersign1\leftbracket#1{\numbersign1}}%
} 
%
% \eqn,\eqnna,eqnnd obsolete pour compatibilite anterieure, 

% 
\def\em@rk{\hbox{}} 
\def\xeqn{\expandafter\xe@n}\def\xe@n(#1){#1}
\def\xeqna#1{\expandafter\xe@na#1}\def\xe@na\hbox#1{\xe@nap #1}
\def\xe@nap$(#1)${\hbox{$#1$}}
% \eqns allows to quote several equations, suppressing unnecessary ()
\def\eqns#1{(\e@ns #1{\hbox{}})}
\def\e@ns#1{\ifx\UNd@FiNeD#1\message{eqnlabel \string#1 is undefined.}%
\xdef#1{(?.?)}\fi{\let\hyperref=\relax\xdef\next{#1}}%
\ifx\next\em@rk\def\next{}%
\else\ifx\next#1\xeqn#1\else\def\n@xt{#1}\ifx\n@xt\next#1\else\xeqna#1\fi
\fi\let\next=\e@ns\fi\next}
%**********************************************************************
%*************************** figure macros ****************************
% Pour centrer ajouter 16mm a la taille de la boite
\def\figure#1#2{\global\advance\nofigure by 1 \vglue#1%
{\elevenpoint
\setbox1=\hbox{#2}
\ifdim\wd1=0pt\centerline{Fig.\ \the\nofigure\hskip0.5mm}%
\else\def\caption{Fig.\ \the\nofigure\quad#2\hskip0mm}
\setbox0=\hbox{\caption}
\ifdim\wd0>\hsize\noindent Fig.\ \the\nofigure\quad#2\else
                 \centerline{\caption}\fi\fi}\par}
% le bigskip a la fin a ete enleve!
 % obsolete, for compatibility
%***************
%figure alignee a gauche
\def\lfigure#1#2{\global\advance\nofigure by
1\vglue#1\leftline{\elevenpoint\hskip10truemm  Fig.\
\the\nofigure\quad #2}} 
%***********************************************************************
\catcode`@=12

\def\draftend{\vfill\supereject%
\immediate\closeout\equa\immediate\closeout\tab
\ifdraftmode%\vfill\supereject%
{\bf \titlename},\par ------------ Date \today. -----------\par
\edef\ewrite{\write\eqdf{}}\ewrite%
%\immediate\closeout\eqdf 
\catcode`\&=0
\catcode`\\=10
\input \equation
\catcode`\\=0
\catcode`\&=4\fi
\end
}

\input epsf
\def\rbook{[\hyperref {}{reference}{1}{1}]}
\def\rParisi{[\hyperref {}{reference}{2}{2}]}
\def\rNMB{[\hyperref {}{reference}{3}{3}]}
\def\rBNGMo{[\hyperref {}{reference}{4}{4}]}
\def\rLGZJi{[\hyperref {}{reference}{5}{5}]}
\def\rLGZJii{[\hyperref {}{reference}{6}{6}]}
\def\reviewLGZJ{[\hyperref {}{reference}{7}{7}]}
\def\repsilonv{[\hyperref {}{reference}{8}{8}]}
\def\rLGZJiii{[\hyperref {}{reference}{9}{9}]}
\def\rIsexpo{[\hyperref {}{reference}{10}{10}]}
\def\rLipa{[\hyperref {}{reference}{11}{11}]}
\def\rNickel{[\hyperref {}{reference}{12}{12}]}
\def\rZJhts{[\hyperref {}{reference}{13}{13}]}
\def\rAdler{[\hyperref {}{reference}{14}{14}]}
\def\rFiCh{[\hyperref {}{reference}{15}{15}]}
\def\rGutt{[\hyperref {}{reference}{16}{16}]}
\def\rNiRe{[\hyperref {}{reference}{17}{17}]}
\def\rBCGS{[\hyperref {}{reference}{18}{18}]}
\def\rBloet{[\hyperref {}{reference}{19}{19}]}
\def\rRaGu{[\hyperref {}{reference}{20}{20}]}
\def\rBWW{[\hyperref {}{reference}{21}{21}]}
\def\rWZAN{[\hyperref {}{reference}{22}{22}]}
\def\rNA{[\hyperref {}{reference}{23}{23}]}
\def\ramprat{[\hyperref {}{reference}{24}{24}]}
\def\rbervil{[\hyperref {}{reference}{25}{25}]}
\def\rBBMN{[\hyperref {}{reference}{26}{26}]}
\def\rHaDo{[\hyperref {}{reference}{27}{27}]}
\def\rRAJA{[\hyperref {}{reference}{28}{28}]}
\def\rBB{[\hyperref {}{reference}{29}{29}]}
\def\rMunster{[\hyperref {}{reference}{30}{30}]}
\def\rJoseph{[\hyperref {}{reference}{31}{31}]}
\def\ssintro{\hyperref {}{section}{1}{1}}
\def\eaction{($\hyperref {}{equation}{1.1}{1.1}$)}
\def\edeftemp{($\hyperref {}{equation}{1.2}{1.2}$)}
\def\ssefact{\hyperref {}{section}{2}{2}}
\def\egrzerom#1{(\relax \hyperref {}{equation}{2.1#1}{\hbox {$2.1#1$}})}
\def\ehscal{($\hyperref {}{equation}{2.3}{2.3}$)}
\def\ezscalvar{($\hyperref {}{equation}{2.4}{2.4}$)}
\def\eVpotFeq{($\hyperref {}{equation}{2.6}{2.6}$)}
\def\ezvaria{($\hyperref {}{equation}{2.8}{2.8}$)}
\def\ehsmallm{($\hyperref {}{equation}{2.9}{2.9}$)}
\def\estaDfix{($\hyperref {}{equation}{2.10}{2.10}$)}
\def\egabede{($\hyperref {}{equation}{2.11}{2.11}$)}
\def\echichi{($\hyperref {}{equation}{2.13}{2.13}$)}
\def\ssepsexp{\hyperref {}{section}{3}{3}}
\def\rBrKr{[\hyperref {}{reference}{32}{32}]}
\def\estateps{($\hyperref {}{equation}{3.1}{3.1}$)}
\def\econstlamb{($\hyperref {}{equation}{3.2}{3.2}$)}
\def\estateii#1{(\relax \hyperref {}{equation}{3.3#1}{\hbox {$3.3#1$}})}
\def\sssUQepst{\hyperref {}{section}{4}{4}}
\def\eJoseph{($\hyperref {}{equation}{4.1}{4.1}$)}
\def\ezmaptheta{($\hyperref {}{equation}{4.2}{4.2}$)}
\def\emagfparii{($\hyperref {}{equation}{4.5}{4.5}$)}
\def\eFHrelat#1{(\relax \hyperref {}{equation}{4.8#1}{\hbox {$4.8#1$}})}
\def\edfGamsg{($\hyperref {}{equation}{4.9}{4.9}$)}
\def\eGamJos{($\hyperref {}{equation}{4.10}{4.10}$)}
\def\esuscJos{($\hyperref {}{equation}{4.12}{4.12}$)}
\def\eratCC{($\hyperref {}{equation}{4.13}{4.13}$)}
\def\eratAA{($\hyperref {}{equation}{4.14}{4.14}$)}
\def\eratRc{($\hyperref {}{equation}{4.15}{4.15}$)}
\def\eratRchi{($\hyperref {}{equation}{4.16}{4.16}$)}
\def\estatJos{($\hyperref {}{equation}{4.17}{4.17}$)}
\def\estatparm#1{(\relax \hyperref {}{equation}{4.18#1}{\hbox {$4.18#1$}})}
\def\estatparm{($\hyperref {}{equation}{4.19}{4.19}$)}
\def\edevbc{($\hyperref {}{equation}{4.20}{4.20}$)}
\def\eCpm#1{(\relax \hyperref {}{equation}{4.22#1}{\hbox {$4.22#1$}})}
\def\rAnSo{[\hyperref {}{reference}{33}{33}]}
\def\ssUQfixD{\hyperref {}{section}{5}{5}}
\def\egNick{($\hyperref {}{equation}{5.4}{5.4}$)}
\def\eoneloop{($\hyperref {}{equation}{5.5}{5.5}$)}
\def\ettozero{($\hyperref {}{equation}{5.6}{5.6}$)}
\def\ezphase{($\hyperref {}{equation}{5.7}{5.7}$)}
\def\rOMD{[\hyperref {}{reference}{34}{34}]}
\def\ehtetv{($\hyperref {}{equation}{5.10}{5.10}$)}
\def\erhomin{($\hyperref {}{equation}{5.11}{5.11}$)}
\def\rBorsom{[\hyperref {}{reference}{35}{35}]}
\def\rLoef{[\hyperref {}{reference}{36}{36}]}
\def\ssUQsum{\hyperref {}{section}{6}{6}}
\def\eseries{($\hyperref {}{equation}{6.1}{6.1}$)}
\def\eBoreltr{($\hyperref {}{equation}{6.2}{6.2}$)}
\def\eaLOB{($\hyperref {}{equation}{6.4}{6.4}$)}
\def\rLGZJed{[\hyperref {}{reference}{37}{37}]}
\def\rPABRE{[\hyperref {}{reference}{38}{38}]}
\def\rZLFish{[\hyperref {}{reference}{39}{39}]}
\def\BLGZJ{[\hyperref {}{reference}{40}{40}]}
\def\rsokolov{[\hyperref {}{reference}{41}{41}]}
\def\rreisz{[\hyperref {}{reference}{42}{42}]}
\def\rbuco{[\hyperref {}{reference}{43}{43}]}
\def\rtsypin{[\hyperref {}{reference}{44}{44}]}
\def\rkimlandau{[\hyperref {}{reference}{45}{45}]}
\def\rwetterich{[\hyperref {}{reference}{46}{46}]}
\def\rmorris{[\hyperref {}{reference}{47}{47}]}
\def\rfishertarkoa{[\hyperref {}{reference}{48}{48}]}
\def\rgauntdomb{[\hyperref {}{reference}{49}{49}]}
\def\rwetterichb{[\hyperref {}{reference}{50}{50}]}
\def\mainresult{($\hyperref {}{equation}{7.1}{7.1}$)}
\def\eWidomnum{($\hyperref {}{equation}{7.3}{7.3}$)}
\def\rliu{[\hyperref {}{reference}{51}{51}]}
\def\raharony{[\hyperref {}{reference}{52}{52}]}
\def\rprivman{[\hyperref {}{reference}{53}{53}]}
\def\ssUQampl{\hyperref {}{section}{8}{8}}
\def\msveff{($\hyperref {}{equation}{A1.1}{A1.1}$)}
\def\msrg{($\hyperref {}{equation}{A1.2}{A1.2}$)}
\def\ermstom{($\hyperref {}{equation}{A1.3}{A1.3}$)}

%\draft

\francmodefalse
\def\r{{\rm r}}

\def\phib{\phi}

\def\F{\tilde F}

\newskip\tableskipamount \tableskipamount=8pt plus 3pt minus 3pt
\def\tableskip{\vskip\tableskipamount}
\def\g{\tilde{g}}
\preprint{SPhT/96-116}

\title{3D Ising model: the scaling equation of state}

\centerline{R.~Guida* and J.~Zinn-Justin**}
\medskip{\it
CEA-Saclay, Service de Physique Th\'eorique***, F-91191 Gif-sur-Yvette
\goodbreak Cedex, FRANCE} 
\footnote{}{${}^{*}$email: guida@spht.saclay.cea.fr}
\footnote{}{${}^{**}$email: zinn@spht.saclay.cea.fr}
\footnote{}{${}^{***}$Laboratoire de la Direction des
Sciences de la Mati\`ere du 
Commissariat \`a l'Energie Atomique}

\abstract
The equation of state of the universality class of the 3D Ising model
is determined numerically in the critical domain from quantum field theory and
renormalization group techniques. The starting point is the five loop
perturbative expansion of the effective potential (or free energy) in the
framework of renormalized $\phi^4_3$ field theory. The 3D perturbative
expansion is summed, using a Borel transformation and a mapping based on large
order behaviour results. It is known that the equation of state has 
parametric representations which incorporate in a simple way its scaling and
regularity properties. We show that such a representation can be used to
accurately determine it from the knowledge of the few first coefficients of
the expansion for small magnetization. Revised values of amplitude ratios are
deduced. Finally we compare the 3D values with the results obtained by the
same method from the $\varepsilon=4-d$ expansion. 
\endabstract
\nref\rbook{J. Zinn-Justin, 1989, {\it Quantum Field
Theory and Critical Phenomena}, in particular chap.~28 of third ed., Clarendon
Press (Oxford 1989, third ed. 1996).}
\nref\rParisi{G. Parisi, {\it Carg\`ese Lectures 1973}, published in
{\it J. Stat. Phys.} 23 (1980) 49.}
\nref\rNMB{B.G. Nickel, D.I.Meiron, G.B. Baker, Univ. of Guelph Report 1977.}
\nref\rBNGMo{G.A. Baker, B.G. Nickel, M.S. Green
and D.I. Meiron, {\it Phys. Rev. Lett.} 36  (1976) 1351.}
\nref\rLGZJi{J.C. Le Guillou and J. Zinn-Justin, {\it
Phys. Rev. Lett.} 39 (1977) 95.} 
\nref\rLGZJii{J.C. Le Guillou and J. Zinn-Justin,{\it Phys. Rev.} B21 (1980)
3976.}  
\nref\reviewLGZJ{{\it Phase
Transitions} vol. B72, M. L\'evy, J.C. Le Guillou and J.~Zinn-Justin eds.
Proceedings of the Carg\`ese Summer Institute 1980 (Plenum, New York
1982).}
\nref\repsilonv{A.A. Vladimirov,
D.I. Kazakov and O.V. Tarasov, {\it Zh. Eksp. Teor. Fiz.} 77 (1979) 1035 ({\it
Sov. Phys. JETP} 50 (1979) 521)\semi 
K.G. Chetyrkin, A.L. Kataev and F.V. Tkachov, {\it Phys. Lett.} B99 (1981)
147; B101 (1981) 457(E)\semi
K.G. Chetyrkin and F.V. Tkachov, {\it Nucl. Phys.} B192 (1981) 159\semi
K.G. Chetyrkin, S.G. Gorishny, S.A. Larin and F.V. Tkachov, {\it Phys. Lett.}
132B (1983) 351\semi
D.I. Kazakov, {\it Phys. Lett.} 133B (1983) 406\semi
S.G. Gorishny, S.A. Larin and F.V. Tkachov, {\it Phys. Lett.} 101A (1984)
120\semi
H. Kleinert, J. Neu, V. Schulte-Frohlinde, K.G. Chetyrkin and S.A.
Larin, {\it Phys. Lett.} B272 (1991) 39, Erratum B319 (1993) 545.}
\nref\rLGZJiii{J.C. Le Guillou
and J. Zinn-Justin, {\it J. Physique Lett. (Paris)} 46 (1985) L137; {\it J.
Physique (Paris)} 48 (1987) 19; 50 (1989) 1365.}
\nref\rIsexpo{A.C. Flewelling, R.J. Defonseka, N. Khaleeli, J. Partee and D.T.
Jacobs, {\it J. Chem. Phys.} 104 (1996) 8048\semi
C.A. Ramos, A.R. King and V. Jaccarino, {\it Phys. Rev. B} 40 (1989) 7124.}
\nref\rLipa{ J.A. Lipa, D.R. Swanson, J. Nissen,
T.C.P. Chui and U.E. Israelson, {\it Phys. Rev. Lett.} 76 (1996) 944.}
\nref\rNickel{B.G. Nickel, {\it Physica} 106A (1981) 48.}
\nref\rZJhts{J. Zinn-Justin, {J. Physique (Paris)} 42 (1981) 783.}
\nref\rAdler{J. Adler, {\it J. Phys.} A16 (1983) 3585.}
\nref\rFiCh{J.H. Chen, M.E. Fisher and B.G. Nickel, {\it Phys. Rev. Lett.}
48 (1982) 630\semi
M.E. Fisher and J.H. Chen, {\it J. Physique (Paris)} 46 (1985) 1645.}
\nref\rGutt{A.J.~Guttmann, {\it J. Phys.} A20 (1987) 1855\semi
A.J.~Guttmann and I.G.~Enting, {\it J. Phys.} A27 (1994) 8007.} 
\nref\rNiRe{B.G. Nickel and J.J. Rehr, {\it J. Stat. Phys.} 61 (1990) 1.}
\nref\rBCGS{G. Bhanot, M.
Creutz,  U. Gl\"assner and K. Schilling, {\it Phys. Rev.} B49 (1994) 12909.} 
\nref\rBloet{H.W.J. Bl\"ote, A. Compagner, 
J.H. Croockewit, Y.T.J.C. Fonk, J.R. Heringa, A. Hoogland, T.S. Smit and A.L.
van Villingen, {\it Physica} A161 (1989) 1.}
\nref\rRaGu{C.F. Baillie, R. Gupta,
K.A. Hawick and G.S. Pawley, {\it Phys. Rev.} B45 (1992) 10438, and references
therein \semi R. Gupta and P. Tamayo, {\it Critical exponents of the 3D Ising 
model}, LA UR-96-93 preprint {\bf cond-mat/9601048}.}
\nref\rBWW{G.M. Avdeeva and A.A. Migdal, {\it JETP Lett.} 16 (1972) 178\semi
E. Br\'ezin,
D.J. Wallace and K.G. Wilson, {\it Phys. Rev. Lett.} 29 (1972) 591; {\it Phys.
Rev.} B7 (1973) 232.} 
\nref\rWZAN{D.J. Wallace and R.P.K. Zia, {\it J.
Phys. C: Solid State Phys.} 7 (1974) 3480.}
\nref\rNA{J.F. Nicoll and P.C. Albright,
{\it Phys. Rev.} B31 (1985) 4576.}
\nref\ramprat{P.G. Watson, {\it J. Phys. C} 2 (1969) 1883\semi
E. Br\'ezin, J.C. Le Guillou and J.
Zinn-Justin, {\it Phys. Lett.} 47A (1974) 285\semi
H.B. Tarko and M.E. Fisher, {\it Phys. Rev. Lett.} 31 (926) 1973; {\it Phys.
Rev.} B11 (1975) 1217\semi
%C. Bervillier, {\it Phys. Rev.} B14 (1976) 4964\semi
A. Aharony and P.C. Hohenberg, {\it Phys. Rev.} B13 (1976) 3081; {\it Physica}
86-88B (1977) 611\semi
Y. Okabe and K. Ideura, {\it Prog. Theor. Phys.} 66 (1981) 1959
%\semi C. Bervillier, {\it Phys. Rev.} B34 (1986) 8141
.}
\nref\rbervil{C. Bervillier, {\it Phys. Rev.} B34 (1986) 8141;}
\nref\rBBMN{C. Bagnuls, C. Bervillier, D.I. Meiron and B.G.
Nickel, {\it Phys. 
Rev.} B35 (1987) 3585.}
\nref\rHaDo{F.J. Halfkann and V. Dohm, {\it Z.
Phys.} B89 (1992) 79.}
\nref\rRAJA{A.K. Rajantie, Preprint Helsinki Un. HU-TFT-96-22, {\bf
hep-ph/9606216}: In this article the expansion is given in analytic form up
to three loops.} 
\nref\rBB{C. Bagnuls and C. Bervillier, {\it Phys. Rev.} B32
(1985) 7209.}
\nref\rMunster{G. M\"unster and J. Heitger, {\it Nucl. Phys.} B424 (1994)
582\semi C. Gutsfeld, J. K\"uster and G. M\"unster, preprint
{\bf cond-mat/9606091}.}
\nref\rJoseph{P. Schofield, {\it Phys. Rev. Lett.} 23 (1969) 109\semi 
P. Schofield, J.D.
Litster and J.T. Ho, {\it Phys. Rev. Lett.} 23 (1969) 1098\semi B.D.
Josephson, {\it J. Phys. C: Solid State Phys.} 2 (1969) 1113.}
{\bf PACS: } 05.70.Jk,64.60.Fr,11.10.Kk,05.50.+q,64.10.+h,11.15.Tk
\par
{\bf Keywords:}  Field Theory, 
Critical phenomena, Ising model, Equation of state,
Amplitude Ratios, Effective potential,
$d=3$, loop expansion, $\varepsilon$ expansion,
Borel summation, Order Dependent Mapping.
\par
{\bf Corresponding Author:} Riccardo Guida;\par
address: CEA-Saclay, Service de Physique Th\'eorique \par
         F-91191 Gif-sur-Yvette Cedex, France\par
email: guida@spht.saclay.cea.fr;\par
tel: 00 33 1 69088116\par
fax: 00 33 1 69088120. 
\vfill\eject
\section Introduction

Renormalization group (RG) theory of second order phase transitions provides a
complete description of all {\it universal}\/ quantities (for a general
reference on this article see for instance \rbook). Among them the most
studied are critical exponents, because they are easier to calculate, and
because they have been used to test RG predictions by comparing
them with other results (experiments, high or low temperature
series expansion, Monte-Carlo simulations). For the $O(N)$ vector model,
calculations are based upon the $(\phib^2)^2$ field theory \sslbl\ssintro
$${\cal H} (\phib)  = \int \left\{ \ud \left[
\partial_\mu \phib(x)\right]^2+\ud \lambda_2 \phib^2(x) +\frac{1}{4!}\lambda_4
[\phib^2(x)]^2 \right\} \d^{d}x\,.  \eqnd\eaction $$
We recall that near the critical temperature $T_c$  $\lambda_2$ is a linear
measure of the temperature. If we denote 
by $\lambda_{2c}$ the value for which the theory becomes massless ($T=T_c$)
then the parameter $t$
$$t=\lambda_2-\lambda_{2c}\propto T-T_c \,,\eqnd\edeftemp $$
characterizes the deviation from the critical temperature.\par
To determine exponents two strategies have been used. One follows
Parisi's suggestion \rParisi~and is based on perturbation series calculated 
directly in three dimensions. Series up to six loops have been generated
\rNMB~which have been summed by summation methods based on a Borel
transformation (first estimates for $N=1$ were reported in \rBNGMo). The more
accurate estimates are obtained after a conformal mapping which takes full
advantage of the large order behaviour analysis \refs{\rLGZJi,\rLGZJii}.
For a review see \refs{\rbook,\reviewLGZJ}. In this article, as a test of the
variant of the summation method we use, we have reexamined their
determination. \par 
Alternatively the $\varepsilon=4-d$ expansion known up to five loop order
\repsilonv~has been summed by similar techniques \rLGZJiii~(note however that
the authors used the slightly erroneous fifth order of Gorishny {\it et al}). 
Both expansions lead to consistent results. Moreover RG values for critical
exponents have now resisted for many years to confrontation with lattice
calculations and experimental determinations. Table 1 displays results
for Ising-like systems ($N=1$) coming from field theory under the form of
summed $d=3$ series and $\varepsilon$-expansion compared to experiments:
binary fluids ({\it a}), 
liquid--vapour transitions ({\it b}) and antiferromagnets ({\it c}), see
\refs{\reviewLGZJ,\rIsexpo} for references. Only recent Helium superfluid 
transition experiments ($N=2$) \rLipa~now provide results 
which are consistent but more accurate than RG estimates. It would therefore
be interesting to examine whether the series could now be extended to 
further decrease the apparent errors in $d=3$ dimensions. \par
\midinsert
$$ \vbox{\elevenpoint\offinterlineskip\tabskip=0pt\halign to \hsize
{& \vrule#\tabskip=0em plus1em & \strut\ # \ 
& \vrule#& \strut #
& \vrule#& \strut #  
& \vrule#& \strut #  
& \vrule#& \strut #  
& \vrule#& \strut #  
&\vrule#\tabskip=0pt\cr
\noalign{\centerline{Table 1} \tableskip}
\noalign{\centerline{\it Critical exponents for Ising-like systems: RG and
experiments.}
\tableskip} \fileth
height2.0pt& \omit&& \omit&& \omit&& \omit&&\omit&& \omit&\cr
&$ \hfill$&&$ \hfill \gamma \hfill$&&$ \hfill \nu \hfill$&&$ \hfill \beta
\hfill$&&$ \hfill \alpha \hfill$&&$ \hfill \theta=\omega \nu \hfill $&\cr
height2.0pt& 
\omit&&\omit&&\omit&&  \omit&& \omit&&\omit&\cr \fileth
height2.0pt& \omit&&\omit&&\omit&& \omit&& \omit&& \omit&\cr &$
\hfill d=3  \hfill$& &$ \hfill 1.2405\pm0.0015 \hfill$&&$ \hfill
0.6300\pm0.0015  
\hfill$&&$ \hfill 0.325\pm0.0015 \hfill$&&$ \hfill 0.110\pm0.0045 \hfill$&&$
\hfill 0.50\pm0.02 \hfill$&\cr 
height2.0pt& \omit&&\omit&&\omit&& \omit&& \omit&& \omit&\cr &$
\hfill \varepsilon-{\rm exp.} \hfill$& &$ \hfill 1.2390\pm0.0025 \hfill$&&$
\hfill 0.6310\pm0.0015  
\hfill$&&$ \hfill 0.327\pm0.0015 \hfill$&&$ \hfill 0.110\pm0.0045 \hfill$&&$
\hfill 0.51\pm0.03 \hfill$&\cr 
height2.0pt& \omit&&\omit&&\omit&& \omit&& \omit&& \omit&\cr &$
\hfill (a) \hfill$& &$ \hfill 1.236\pm0.008 \hfill$&&$ \hfill 0.625\pm0.010 
\hfill$&&$ \hfill 0.325\pm0.005 \hfill$&&$ \hfill 0.112\pm0.005 \hfill$&&$
\hfill 0.50\pm0.03 \hfill$&\cr 
height2.0pt& \omit&& \omit&&\omit&& \omit&&
\omit&& \omit&\cr&$ \hfill (b) \hfill$& &$ \hfill 1.23\hbox{--}\, 1.25
\hfill$&&$ \hfill 0.625\pm0.006 \hfill$&&$ \hfill 0.316\hbox{--}\, 0.327
\hfill$&&$ \hfill 0.107\pm0.006 \hfill$&&$ \hfill 0.50\pm0.03
\hfill$&\cr 
height2.0pt& \omit&& \omit&&\omit&& \omit&&
\omit&& \omit&\cr&$ \hfill (c) \hfill$& &$ \hfill 1.25\pm 0.01
\hfill$&&$ \hfill 0.64\pm0.01 \hfill$&&$ \hfill 0.328\pm 0.009
\hfill$&&$ \hfill 0.112\pm0.007 \hfill$&&$ \hfill 
\hfill$&\cr 
height2.0pt&\omit&&\omit&& \omit&& \omit&& \omit&& \omit&\cr
\fileth}}$$
\endinsert
Other available estimates come from the analysis of High Temperature series
in lattice models (table 2) and Monte-Carlo simulations. 
For the latter case let us quote two typical set of results in \rBloet~the
values $\nu=0.629\pm0.003$ and $\eta=0.027\pm0.005$ are reported. In
\rRaGu~one finds $\nu=0.625\pm0.001$ and $\eta=.025\pm0.006$ while $\theta$
varies in the range $0.44$--$0.53$.
\midinsert
$$ \vbox{\elevenpoint\offinterlineskip\tabskip=0pt\halign to \hsize
{& \vrule#\tabskip=0em plus1em & \strut\ # \ 
%& \vrule#& \strut #
& \vrule#& \strut #  
& \vrule#& \strut #  
& \vrule#& \strut #  
& \vrule#& \strut #  
&\vrule#\tabskip=0pt\cr
\noalign{\centerline{Table 2} \tableskip}
\noalign{\centerline{\it Critical exponents for Ising-like systems: HT
series.} 
\tableskip} \fileth
height2.0pt& \omit&& \omit&& \omit&&\omit&& \omit&\cr
&$ \hfill$&&$ \hfill \gamma \hfill$&&$ \hfill \nu \hfill$&&$ \hfill \alpha
\hfill$&&$ \hfill \theta=\omega \nu \hfill $&\cr 
height2.0pt& 
\omit&&\omit&&\omit&& \omit&&\omit&\cr \fileth
height2.0pt& \omit&&\omit&&\omit&& \omit&& \omit&\cr &
\hfill \rNickel \hfill& &$ \hfill 1.239\pm0.002 \hfill$&&$ \hfill
0.631\pm0.003  
\hfill$&&$ \hfill \hfill$&&$ \hfill \hfill$&\cr 
height2.0pt& \omit&&\omit&& \omit&& \omit&& \omit&\cr &
\hfill \rZJhts \hfill& &$ \hfill 1.2385\pm0.0025 \hfill$&&$
\hfill 0.6305\pm0.0015  
\hfill$&&$ \hfill \hfill$&&$
\hfill 0.57\pm0.07 \hfill$&\cr 
height2.0pt& \omit&&\omit&& \omit&& \omit&& \omit&\cr &
\hfill \rAdler \hfill& &$ \hfill 1.239\pm0.003 \hfill$&&$ \hfill
0.631\pm0.004  
\hfill$&&$ \hfill  \hfill$&&$ \hfill  \hfill$&\cr 
height2.0pt& \omit&&\omit&& \omit&& \omit&& \omit&\cr &
\hfill \rFiCh \hfill& &$ \hfill 1.2395\pm0.0004 \hfill$&&$ \hfill
0.632\pm0.001  
\hfill$&&$ \hfill 0.105\pm0.007 \hfill$&&$
\hfill 0.54 \pm0.05 \hfill$&\cr 
height2.0pt& \omit&&\omit&& \omit&& \omit&& \omit&\cr &
\hfill \rGutt  \hfill& &$ \hfill 1.239\pm0.003 \hfill$&&$ \hfill
0.632\pm0.003  
\hfill$&&$ \hfill 0.101\pm.004 \hfill$&&$ \hfill  \hfill$&\cr 
height2.0pt& \omit&&\omit&& \omit&& \omit&& \omit&\cr &
\hfill \rNiRe\hfill& &$ \hfill 1.237\pm0.002 \hfill$&&$ \hfill
0.630\pm0.0015  \hfill$&&$ \hfill  \hfill$&&$
\hfill 0.52\pm0.03 \hfill$&\cr 
height2.0pt&\omit&&\omit&& \omit&&
\omit&& \omit&\cr& \hfill \rBCGS \hfill& &$ \hfill 
\hfill$&&$ \hfill  \hfill$&&$ \hfill 0.104\pm0.004 \hfill$&&$
\hfill  \hfill$&\cr 
height2.0pt&\omit&& \omit&& \omit&& \omit&& \omit&\cr
\fileth}}$$
\endinsert
Other universal quantities, like the equation of state 
\refs{\rBWW,\rWZAN,\rNA},
and some amplitude ratios \refs{\ramprat,\rNA,\rbervil}, have also been
calculated but the 
estimates are less accurate because the series are shorter. Perturbative
calculations are more difficult, in particular for $N\ne1$ (continuous
symmetries), due to the presence of two different masses
(transverse and longitudinal) in an external magnetic field and Goldstone
singularities on the coexistence curve ($H=0$, $T<T_c$). Probably also less
effort have been invested up to now. \par 
In this article we therefore present a determination of the equation of state 
for the $N=1$, $d=3$ case, where longer series are available. 
Our calculations
are based on perturbative expansion at fixed $d=3$ dimension \rParisi.
Five loop series for the renormalized
effective potential of the $\phi^4_3$ 
theory have been first reported by Bagnuls {\it et al.} \rBBMN,
but the printed tables contain some serious misprints. These  have been
noticed by Halfkann and Dohm who have published corrected values \rHaDo~(for
detailed explanations see our appendix). Finally full analytic results up to
three loops have recently become available \rRAJA.
\par 
The main technical difficulty that one faces in $d=3$ calculations is how to
continuously extrapolate field theory results from $T>T_c$ to $T<T_c$. Indeed 
because the massless (or critical) theory is IR divergent in perturbation
theory at any fixed dimension $d<4$, calculations can be performed only in the
massive phase (in contrast with the $\varepsilon$-expansion). In \rBB~one
method is  suggested which has then been used 
\rMunster~ (see also \rBBMN)
to predict some amplitude ratios. We present in this article a
different approach that extends a
suggestion in \rbook, and is motivated by the simplicity of the
parametric representation \rJoseph~of the equation of state within the
framework of the $\varepsilon$-expansion \refs{\rWZAN,\rNA}. 
\par
The first step of our approach is a summation of the available perturbative
series to obtain a non-perturbative determination of small field expansion 
of the effective potential (coefficients of $\phi^6,\phi^8,\phi^{10}$) 
of the continuous $\phi^4_3$ field theory at the IR fixed point.
This result is interesting in itself in view  of the recent effort
devoted to the  problem (Monte Carlo lattice simulations, High Temperature 
series and approximate numerical solutions of Exact Renormalization Group).
With the help of a parametric representation we are then able to reconstruct
the full scaling equation of state, which is therefore the main result of our
article. We then use it to calculate several amplitude ratios which have been
considered before in the literature.  
We compare our predictions with other available results 
($\varepsilon$-expansion, High Temperature Expansion, Monte  Carlo,
Exact Renormalization Group and of course experiments).
\par
The set-up of our article is thus the following. In section 2 we briefly
review the properties of the equation of state and the definitions of
amplitude ratios. In section 3 we recall known results about the
$\varepsilon$-expansion, while in section 4 the idea of parametric equation of
state and its 
application to the $\varepsilon$-expansion are presented. In section 5 we
explain the method we have used in our $d=3$ calculations. 
In section 6 the summation of the perturbative expansions for the effective
potential is discussed. The  results concerning the effective potential and
the (parametric) equation of state are reported in section 7, while 
section 8 contains our results for amplitude ratios and some 
concluding remarks.
\section Effective action and equation of state 

In this article the general framework is the massive theory renormalized
at zero momentum. The correlation functions $\Gamma^{(n)}_\r$ of the
renormalized field $\phib_\r =\phib/\sqrt{Z}$ are
fixed by the normalization conditions \sslbl\ssefact
\eqna\egrzerom
$$ \eqalignno{ \Gamma^{(2)}_\r (p;m ,g) & = m^2 +p^2 + O \left(p^4 \right), &
\egrzerom{a} \cr \Gamma^{(4)}_\r \left(p_i=0;m ,g \right) & = m^{4-d} g\,. &
\egrzerom{b}  \cr} $$
In this renormalization scheme one trades the initial parameters
$\lambda_2$ and $\lambda_4$ for $g$ and $m$. The renormalized coupling
constant $g$ is dimensionless. It has to be set to its IR fixed point value
$g^*$. In \rLGZJii~the value $g^*=23.73\pm 0.09 $ has been numerically
estimated from the series published in \rNMB. The mass parameter $m$ is
proportional to the 
physical mass, or inverse correlation length, of the high temperature phase.
It behaves for $t\propto T-T_c\to 0_+$ as $m\propto t^\nu$, where $\nu$ is the
correlation length exponent (see \rbook~for details). 
Therefore the parameter $m$ is singular at
$T_c$ and thus the extrapolation to the low temperature phase is
non-trivial. Within the framework of the $\varepsilon$-expansion instead, it
is possible to first construct the massless theory and then the theory in the
full critical domain as an expansion in powers of the deviation $t$ from the
critical temperature. Equivalently the problem can be solved because scaling
laws are exactly 
satisfied. In the summed $d=3$ perturbative expansion, at the IR fixed point
$g^*$, scaling relations are satisfied only within summation errors.
 The parametric
representation of the equation of state \rJoseph~will provide a
solution to this problem.  
\par
From the conditions \egrzerom{} it follows that the free energy $\cal F$ per
unit volume expressed in terms of the ``renormalized" magnetization $\varphi$,
i.e.~the expectation value of the renormalized field $\phi_\r =\phi/\sqrt{Z}$,
has a small $\varphi$ expansion of the form (in $d$ dimensions) 
$${\cal F}(\varphi)=\Gamma(\varphi)/{\rm vol.}={\cal F}(0)+\ud m^2
\varphi^2+\frac{1}{4!}m^{4-d} g \varphi^4 +O\left(\varphi^6\right), \eqnn $$
where $\Gamma(\varphi)$ is also the generating functional of One Particle
Irreducible correlators restricted to constant
fields, or the effective potential of the renormalized theory.\par 
The equation of state is the relation between the
magnetic field $H$, the magnetization $M=\left<\phib\right>$ (the ``bare"
field expectation value) and the temperature which is represented by the
parameter $t$ (eq.~\edeftemp)
$$H={\del {\cal F}\over\del M}\,.$$
Near the critical point the equation of state has Widom's scaling form
$$H(M,t)=M^\delta f(t/M^{1/\beta}). \eqnd\ehscal $$ 
It is convenient to introduce the rescaled variable $z$
$$\varphi=m^{(d-2)/2}z/\sqrt{g}\,,\eqnd\ezscalvar $$
and set
$${\cal F}(\varphi)-{\cal F}(0)={m^d\over g}V(z,g). \eqnn $$
The  critical behaviour is described 
by the IR fixed point, 
but we keep here the notation $g$ to remind that 
we must first sum up perturbative expansion for $V$ and then
take $g=g^*$. 
\medskip
The equation of state is obtained from the derivative $F$ of the reduced
effective 
potential $V$ with respect to $z$ 
$$F(z,g)={\partial V(z,g)\over \partial z}\,.\eqnd\eVpotFeq $$
Ising-like symmetry implies that $F$ has an expansion of the form
$$F(z,g)=z+\frac{1}{6}z^3+\sum_{l=2}F_{2l+1}(g) z^{2l+1}\,.\eqnn $$
\par
From eqs.~\eqns{\ezscalvar,\ehscal} we conclude
$$z\propto M t^{-\beta}.\eqnd\ezvaria $$
Comparing then the coefficient of $M$ in the small $M$ expansion of $H(M,t)$ 
(see also eq.~\echichi)
$$H(M,t)=H_0 t^\gamma M+O\left(M^3\right) , \eqnd\ehsmallm $$
where $H_0$ is a numerical constant, with the small $z$ expansion of the
function $F(z)$ we conclude
$$H\propto t^{\beta\delta}F(z) ,\eqnd\estaDfix $$
where the relation between exponents
$$\gamma=\beta(\delta-1),\eqnd\egabede $$
has been used.
One property of the function $H(M,t)$ which plays an essential 
role in our analysis
is {\it Griffith's analyticity}: 
it is regular at $t=0$ for $M>0$ fixed, and simultaneously it is regular at
$M=0$  for $t>0$ fixed. 
\medskip
{\it Amplitude ratios.} 
Universal amplitude ratios are numbers characterizing the behaviour of 
thermodynamical quantities near $T_c$ in the critical domain, which in
addition do not depend on the normalizations of magnetic field, magnetization
and temperature. Several amplitude ratios commonly considered in the
literature can be derived from the scaling equation of state: 
\smallskip
{\it The specific heat.} The singular part of the specific heat,
i.e.~the $\phib^2$ 2-point correlation function at zero momentum, behaves
like 
$$C_H=A^{\pm}\left\vert t \right\vert^{-\alpha},\qquad t\propto T-T_c
\rightarrow \pm 0\,. \eqnn $$
The ratio $A^+/A^-$ then is universal.\par
{\it The magnetic susceptibility.} The magnetic susceptibility $\chi$ in
zero field, i.e.~the $\phi$ 2-point function at zero momentum, diverges like 
$$\chi = C^{\pm}\left\vert t \right\vert^{-\gamma}, \qquad t \rightarrow \pm 0
\, . \eqnd\echichi $$
The ratio $C^+/C^-$ then is also universal.\par
{\it Other amplitude ratios.} 
On the critical isotherm the magnetic susceptibility behaves as
$$\chi = C^c/H^{1-1/\delta}; \eqnn $$
the spontaneous magnetization vanishes as
$$M=B\left(-t\right)^{\beta}.\eqnn $$
One can then define the following universal ratio:
$$ R_c  =\alpha A^+C^+/B^2 ,\eqnn $$
which corresponds to the relation between exponents
$$\alpha+ 2\beta + \gamma =2\, .$$
Indeed using this relation we verify that $R_c$ is proportional to
$F(0,t)M^{-2}\chi $ which is normalization independent.
Aharony and Hohenberg 
define the following universal combination
$$R_\chi=B^{\delta-1}C^+/(C^c \delta)^\delta \,,$$ 
which corresponds to the relation \egabede. It is related
to the quantity $Q_1$ defined by Fisher and Tarko, $R_\chi= Q_1^{-\delta}$.
\section The $\varepsilon$-expansion

Let us first recall the results concerning exponents and the equation of state
which  have been obtained within the framework of the $\varepsilon=4-d$
expansion.  \sslbl\ssepsexp
\subsection Critical exponents

Although the RG functions of the $\phi^4$ theory and therefore the critical
exponents are known up to five-loop order \repsilonv, we give the expansions
only up to the order needed in this article, i.e.~to three loops, referring to
the literature for details. The zero $g^*(\varepsilon)$ of the
$\beta$-function is $ g^*(\varepsilon) =16\pi^2\varepsilon/3
+O(\varepsilon^2)$. 
The values of the critical exponents $\gamma$, $\beta$ and
$\delta=1+\gamma/\beta$ are
$$\eqalign{\gamma&=1+\frac{1}{6}\varepsilon+\frac{25}{324}\varepsilon^2+
\left(\frac{701}{17496}-\frac{2}{27}\zeta(3)\right)\varepsilon^3
+O\left(\varepsilon^4\right) \cr
2\beta&=1-\frac{1}{3}\varepsilon+\frac{1}{81}\varepsilon^2
+\left(\frac{163}{8748}-\frac{2}{27}\zeta(3)\right)\varepsilon^3+
+O\left(\varepsilon^4\right)\cr
\delta&=3\left(1+\frac{1}{3}\varepsilon+\frac{25}{162}\varepsilon^2+
\frac{539}{8748}\varepsilon^3\right)+O\left(\varepsilon^4\right) ,\cr}$$ 
in which $\zeta(s)$ is the Riemann $\zeta$-function. (At orders four and five
$\zeta(5)$ and $\zeta(7)$ successively appear. It has also been shown that
starting at six loop order new non-$\zeta$ like numbers will appear
\ref\rBrKr{D.J. Broadhurst and D. Kreimer, {\it Int. J. Mod. Phys.} C6 (1995)
519.}.)  Other exponents can be obtained from scaling relations. 
\subsection The scaling equation of state

The $\varepsilon$-expansion of the scaling equation of state (eq.~\ehscal)
has been determined up to order $\varepsilon^2$ for the general $O(N)$ model,
and order $\varepsilon^3$ for $N=1$. 
Adjusting the normalizations of $H$ and $M$ to simplify the analytic
expressions one finds at order $\varepsilon^2$:
$$f(x)=1+x+\varepsilon f_1(x)+\varepsilon^2 f_2(x)+\varepsilon^3 f_3(x)+
O\left(\varepsilon^4\right)  ,\eqnd\estateps $$
with:
%\eqna\estate
$$\eqalign{ f_1(x) & = \frac{1}{6}(x+3) L  \cr
f_2(x) &= \frac{1}{72}(x+9)L^2+\frac{25}{324}(x+3)
L \cr
f_3(x)&= {{{L^3}\left( 27 + x \right) }\over {1296}} + 
    {{{L^2}\left( 675 + 246x + 25{x^2} \right) }\over 
      {1944\left( 3 + x \right) }}
 \cr
 &
+ 
    {{L\left( 1617 + 701x - 1296x{\zeta}(3) \right) }\over {17496}}
+{{{x^2}\left( -1 - 2\lambda  + 4{\zeta}(3) \right) }\over 
      {108\left( 3 + x \right) }} 
}
$$
with 
$$L=\log(x+3),$$
while $\lambda$ is a constant
$$\lambda  =\frac{1}{3}\psi'(1/3)-\frac{2}{9}\pi^2=1.17195361934 \ldots\
. \eqnd\econstlamb $$ 
The expression \estateps~is not valid for $x$ large, i.e.~for small
magnetization $M$. In this regime the magnetic field $H$ has a regular
expansion in odd powers of $M$, i.e.~in the variable $z \propto x^{-\beta}$.
Substituting in eq.~\estateps~$x=x_0 z^{-1/\beta}$ (the constant $x_0$
takes care of the normalization of $z$) and expanding in
$\varepsilon$  one finds at order $\varepsilon^3$ for the function \eVpotFeq
$$F(z)=\F_0(z)+\varepsilon \F_1(z)+\varepsilon^2 \F_2(z)+
\varepsilon^3 \F_3(z),$$
with
\def\L{\tilde L}
\eqna\estateii
$$ \eqalignno{\F_0& = z + \frac{1}{6}z^3 &\estateii{a}\cr
\F_1& =\frac{1}{12}\bigl(-z^3 + \L( 2z + z^3 )\bigr)\cr
\F_2& = \frac{1}{1296}\bigl(-50z^3 + \L( 100z - 4z^3 ) +\L^2( 18z +
27z^3)\bigr) &\estateii{b}\cr 
\F_3& = \frac{1}{69984}{1\over ( 2 + z^2 )} 
\left[\L^3( 108z + 540z^3 + 243z^5) +\L^2( 1800z + 1494z^3 + 621z^5)\right.\cr
&\quad+ \L\bigl(-1622 z^5  + z^3(1504 - 5184\zeta(3)) + z( 5608 -
10368\zeta(3) )\bigr) \cr
&\quad\left. +  z^5\bigl(-997 - 648\lambda + 3888\zeta(3)\bigr)
+z^3\bigl(-2804 + 5184\zeta(3)\bigr)\right]&\estateii{c}\cr}$$
and
$$\L=\log(1+z^2/2).$$

While within the framework of the formal $\varepsilon$-expansion one can
easily pass from one expansion to the other one, still a matching problem
arises if one wants to apply the $\varepsilon$-expansion for $d=3$,
i.e.~$\varepsilon=1$. One is thus naturally led to look for a uniform
representation of the equation of state valid in both limits. 
Josephson-Schofield
parametric representation has this property. 
\section Parametric representation of the equation of state

We parametrize $M$ and $t$ in terms of two variables $R$ and $\theta$,
setting\sslbl\sssUQepst: 
$$\left\lbrace\eqalign{M &= m_0 R^{\beta}\theta\, , \cr t& = R
\left(1-\theta^2 \right),  \cr H & =h_0 R^{\beta\delta}h(\theta)\, ,
\cr}\right. \eqnd\eJoseph $$
where $h_0, m_0$ are two normalization constants~\rJoseph. This
parametrization also
corresponds in terms of the scaling variables $x$ of eq.~\estateps{} or $z$
from eq.~\ezscalvar~to set 
$$\eqalignno{z&=\rho \theta/ (1-\theta^2)^\beta,\quad \theta >
0\,, & \eqnd\ezmaptheta \cr
x&=x_0\rho^{-1/\beta}\left(1-\theta^2\right)\theta^{-1/\beta} ,&\eqnn \cr
}$$
where $\rho$ is some other positive constant. \par 
Then the function $h(\theta)$   
%$$h(\theta)=\theta^{\delta}f\left(x(\theta)\right)/h_0, \eqnd\emagfpar $$
is an odd function of $\theta$ regular near $\theta =1$, which is $x$ small,
and near $\theta=0$ which is $x$ large. We choose $h_0$ such that
$$ h(\theta)=\theta+O\left(\theta^3\right)\, .$$
The function $h(\theta)$ vanishes for $\theta=\theta_0$ which corresponds to
the coexistence curve $H=0,T<T_c$. From $\theta_0$ we obtain the universal
rescaled spontaneous magnetization $|z_0|$
$$|z_0|=\rho \theta_0/ (\theta_0^2-1)^\beta.  \eqnn $$
Note that the mapping \ezmaptheta~is not invertible for values of $\theta$
such that $z'(\theta)=0$. One verifies that the derivative vanishes for
$\theta^2=1/(1-2\beta)\sim 2.86$. We shall see that this value is reasonably
larger than the largest value of $\theta^2=\theta_0^2$ we will have to
consider. 
\par
Finally it is useful for later purpose to write more explicitly the relation
between the function $F(z)$ of eq.~\eVpotFeq~and the function $h(\theta)$:
$$h(\theta)=\rho^{-1}
\left(1-\theta^2\right)^{\beta\delta}F\left(z(\theta)\right). \eqnd\emagfparii
$$ 
Expanding both functions
$$\eqalignno{F(z)&=z+\frac{1}{6}z^3+\sum_{l=2}F_{2l+1}z^{2l+1},&\eqnn \cr
h(\theta)/\theta&=1+\sum_{l=1}h_{2l+1} \theta^{2l} ,&\eqnn \cr}$$
we find the relations
\eqna\eFHrelat
$$\eqalignno{h_3&=\frac{1}{6}\rho^2-\gamma & \eFHrelat{a}\cr
h_5&=\ud\gamma(\gamma-1)+\frac{1}{6}(2\beta-\gamma)\rho^2+F_5\rho^4 & 
\eFHrelat{b}\cr
h_7&=\frac{1}{6}\gamma(\gamma-1)(\gamma-2)+\frac{1}{12}(2\beta-\gamma)
(2\beta-\gamma+1)\rho^2\cr&\quad +(4\beta-\gamma)F_5\rho^4 +F_7\rho^6
&\eFHrelat{c} 
\cr &\cdots \cr}$$
From the parametric representation of the equation of state it is also
possible to derive a representation for the singular part of the free energy
per unit volume. Setting:
$${\cal F}_{\rm sg.}(M,t)\equiv \Gamma_{\rm sg.}(M,t)/{\rm vol.}
= h_0 m_0 R^{2-\alpha}g(\theta),
\eqnd\edfGamsg $$ 
one finds for $g(\theta)$  a differential equation:
$$h(\theta)\left(1-\theta^2+2\beta\theta^2\right)=2(2-\alpha)\theta
g(\theta)+ \left(1-\theta^2\right) g'(\theta). \eqnd{\eGamJos} $$
The integration constant is fixed by requiring the regularity of
$g(\theta)$ at $\theta=1$.
In the same way the inverse magnetic susceptibility is given by:
$$\chi^{-1}= (h_0/m_0) R^{\gamma}g_2(\theta), \eqnn $$
with:
$$g_2(\theta)\left(1-\theta^2+2\beta\theta^2\right)=2\beta\delta\theta
h(\theta)+ \left(1-\theta^2\right) h'(\theta). \eqnd{\esuscJos} $$
Note that $g_2$ then in general has a pole at $\theta^2=1/(1-2\beta)$,
the point at which the mapping \ezmaptheta~is not invertible.\par
The functions $g(\theta)$ and $g_2(\theta)$ will be used to calculate several
universal ratios of amplitudes.
\subsection Amplitude ratios

The amplitude ratios defined in section \ssefact~are related to the scaling
equation of state. They can be calculated from the functions $h(\theta)$,
$g(\theta)$ and $g_2(\theta)$ appearing in the parametric representation, and
thus ultimately from the function $h(\theta)$ alone
(eqs.~\eqns{\eGamJos,\esuscJos}). One verifies that indeed they do not depend 
on the variable $R$ and the constants $m_0,h_0$ appearing in \eJoseph.
\smallskip
{\it The magnetic susceptibility.} The magnetic susceptibility in zero field
can be calculated from the function $g_2(\theta)$ defined by equation
\esuscJos. One obtains (equation \esuscJos):
$$ {C^+ \over C^-} =(\theta_0^2-1)^{-\gamma}
{ h'(\theta_0)(1-\theta_0^2)\over h'(0)(1-\theta_0^2+2\beta
\theta_0^2)} .\eqnd\eratCC $$
\medskip
{\it The specific heat.} This ratio is directly related to the
function $g(\theta)$ defined by equation \eGamJos:
$${A^+ \over A^-}= \left(\theta_0^2-1\right)^{2-\alpha}{g(0) \over
g(\theta_0)} . \eqnd\eratAA $$  
\medskip
{\it Other universal ratios.} 
Similarly the ratios $R_c$ and $R_\chi$ can be derived from the functions
$g(\theta)$ and $h(\theta)$
$$\eqalignno{R_c &
=-\alpha(1-\alpha)(2-\alpha){g(0)(\theta_0^2-1)^{2\beta}\over 
h'(0)\theta_0^2}, & \eqnd\eratRc \cr   
R_\chi& ={h(1)\theta_0^{\delta-1} \over h'(0) (\theta_0^2-1)^\gamma}. &
\eqnd\eratRchi 
\cr} $$
% why the minus sign in R_c ?
%
\subsection Parametric representation in the $\varepsilon$-expansion

Up to order $\varepsilon^2$ the constant $m_0$ (or $\rho$) can be chosen in
such a way that the function $h(\theta)$ reduces to:  
$$ h(\theta)= \theta\left(1 -\frac{2}{3}\theta^2 \right) +O\left(\varepsilon^2
\right). \eqnd{\estatJos} $$
The simple model in which $h(\theta)$ is approximated by a cubic odd function
of $\theta$ is called the linear parametric model. At order $\varepsilon^2$
the linear parametric model is exact, but at order $\varepsilon^3$ the
introduction of a term proportional to $\theta^5$ becomes necessary \refs{\rWZAN,\rNA}.
One finds:   
\eqna\estatparm
%$$\eqalignno{h(\theta)&=\theta\left(1-\theta^2/b^2\right)\left(1+c\theta^2
%\right)+O\left( \varepsilon^4\right)&\estatparm{a} \cr
%&=\theta (1+h_3\theta^2+h_5\theta^4)+O\left( \varepsilon^4\right),
%&\estatparm{b}  \cr} $$
$$h(\theta)=\theta (1+h_3\theta^2+h_5\theta^4)+O\left(\varepsilon^4\right),
\eqnd \estatparm $$ 
with
$$h_3=-{2\over3}\left(1+{\varepsilon^2 \over 12} \right) ,\quad
h_5={\varepsilon^3 \over 27}\left(\zeta(3)-\ud\lambda -
\frac{1}{4}\right),\eqnd\edevbc $$  
where $\lambda$ is the constant \econstlamb.\par
The function $h(\theta)$ vanishes on the coexistence curve for
$\theta=\theta_0$: 
$$\theta_0^2  ={3 \over 2}\left(1-{\varepsilon^2 \over 12} \right)
+O(\varepsilon^3) .
\eqnn $$ 
Note that $h_3$ and thus $\theta_0$  are determined only up to order
$\varepsilon^2$. 
It follows
$$\rho^2=6(\gamma+h_3)=2\left(1+\ud\varepsilon+\frac{7}{108}\varepsilon^2
\right)=3.13\pm0.13,$$ 
because $h_3$ is determined only up to order $\varepsilon^2$, and
$$\eqalign{|z_0|&= \rho\theta_0(\theta_0^2-1)^{-\beta} \cr
&=\sqrt{3}\times 2^\beta
\left[1+\frac{1}{4}\varepsilon+\frac{73}{864}\varepsilon^2 
+\left(\frac{1}{24}\lambda-\frac{7}{36}\zeta(3)+\frac{5581}{93312}\right)
\varepsilon^3\right]\sim 2.87\pm0.06\ \cr}  $$
(in this case we summed by a Pad\'e $[1,2]$).
\medskip
{\it Remark.} Even in the more general $O(N)$ case, the  
parametric representation automatically satisfies the different
requirements about the regularity properties of the equation of state and
leads to uniform approximations. However for $N>1$ the function $h(\theta)$
still has a singularity on the coexistence curve, due to the presence of
Goldstone modes in the ordered phase and has therefore a more complicated
form. The nature of this singularity can be obtained from the study of the
non-linear $\sigma$-model. It is not clear whether a simple polynomial
approximation would be useful. For $N=1$ instead, one expects at most
an essential singularity on the coexistence curve, due to barrier penetration,
which is much weaker and non-perturbative in the small $\varepsilon$- or small
$g$-expansion.
\medskip
{\it Amplitude ratios.} The use of eqs.~\eqns{\eratAA{--}\eratRchi}
together with the expression of $h(\theta)$
eqs.~\eqns{\estatparm,\edevbc}
evaluated at 
$\varepsilon =1$ 
gives us the predictions for the amplitude ratios reported in table 5
($\varepsilon$-expansion $(b)$). 
Moreover from these equations
the known $\varepsilon$-expansion of various amplitude ratios,
\refs{\ramprat,\rNA,\rbervil},
 can be 
easily 
obtained. The ratio $C^+ / C^-$, related to the magnetic susceptibility in
zero field, is
\eqna\eCpm
$$\eqalignno{ {C^+ \over C^-} & = {2^{\gamma+1} \over 6\beta-1}\left[ 1+
\left( 
{2\lambda+1 \over 4}-\zeta(3)\right){\varepsilon^3 \over 12}\right]
+O\left(\varepsilon^4 \right)  & \eCpm{a} \cr
&=2^{\gamma-1}(\delta-1)\left[1+\frac{1}{36}\left(
\zeta(3)+\frac{3}{2}\lambda+\frac{1}{4}\right)\varepsilon^3\right]&
\eCpm{b}\cr 
&=2^\gamma\left[1+\ud\varepsilon+\frac{25}{108}\varepsilon^2+
\left(\frac{1159}{11664}+\frac{1}{36}\zeta(3)
+\frac{1}{24}\lambda\right)\varepsilon^3\right]. & \eCpm{c}  \cr} $$
The ratio $C^+ / C^-$ can be expressed at order $\varepsilon^2$ entirely in
terms of critical exponents. This form follows naturally from the parametric
representation of the equation of state. The $\varepsilon^3$ relative
correction is  of the order of only 3\%. The three first  expressions yield
for $\varepsilon=1$ respectively (the exponents being replaced by the central
values of the summed $\varepsilon$-expansion): 
$$ {C^+ \over C^-}=4.688,\ 4.757,\ 4.863\ .$$
The spread gives for this short series an indication about the uncertainty
about the result.\par
The ratio $A^+/A^-$ at order $\varepsilon^2$ is given by
$${A^+ \over A^-}=2^{\alpha-2}\left[1+\varepsilon+\left(\frac{43}{54}
-\frac{1}{6}\lambda-\zeta(3)\right)\varepsilon^2\right]
+O(\varepsilon^3).$$
The $\varepsilon$-expansion of $R_c$ is:
$$R_c = {1 \over 9}2^{-2\beta-1}\varepsilon
\left[1+\frac{17}{27}\varepsilon+\left(\frac{989}{2916}-\frac{4}{9}\zeta(3)
-\frac{2}{9}\lambda\right)\varepsilon^2\right] +O\left(\varepsilon^4\right).
\eqnn $$    
$R_{\chi}$ is given by
%$$R_{\chi}=Q_1^{-\delta}=(2-\alpha)(1-\alpha){g(0)b^{\delta-1}\over
%h(1)(b^2-1)^\gamma} .$$
%$$Q_1 =3\left({2^{1-2\beta}\over 27}\right)^{(\delta-1)/2\delta}
%\left[ 1+ \left( \zeta(3)-{2\lambda+1 \over 4}\right) {\varepsilon^3 \over
%54} \right] +O\left(\varepsilon^4\right). \eqnn $$ 
$$R_\chi=3^{(\delta-3)/2}2^{\gamma+(1-\delta)/2}\left[ 1+\left(\frac{1}{72}+
\frac{1}{36}\lambda-\frac{1}{18}\zeta(3)\right) \varepsilon^3 \right]
+O\left(\varepsilon^4\right). \eqnn $$  

\nref\rAnSo{S.A. Antonenko and A.I. Sokolov, {\it Phys. Rev.} E51 (1995) 1894:
in this article the expressions for the general $O(N)$ theory are reported.}
\section The perturbative expansion at fixed dimension three

Critical exponents and several other universal quantities have been estimated
using the perturbative expansion at fixed dimension $d<4$. 
Since the massless theory is then IR divergent, calculations have been
performed within the framework of the 3D perturbative expansion renormalized
at zero momentum for the massive $\phi^4$ field theory.
As in the case of the $\varepsilon$-expansion, it is necessary to  first
determine the IR stable zero $g^*$ of the function $\beta(g)$  which is given
by a few terms of a  divergent expansion. The obvious problem is that we have
no longer a small parameter in which to expand and $g^*$ is a number of order
1. Already at this point a summation method is required. Note also that at any
finite order the results for universal quantities become renormalization
scheme dependent in contrast with the results of the
$\varepsilon$-expansion.\sslbl\ssUQfixD \par  
On the other hand, because one-loop diagrams have, in 3 dimensions, a simple
analytic expression, it has been possible to extend the calculation of RG
functions in the $N$-vector model up to six-loop order. The expansions
are \refs{\rNMB,\rAnSo}:   
$$\eqalignno{ \beta(\g) & =-\g+ \g^2 -\frac{308}{729} \g^3 + 0.3510695978\g^4
-0.3765268283 \g^5 & \cr & \quad + 0.49554751 \g^6 - 0.749689 \g^7
+O\left(\g^8\right),& \eqnn \cr
\gamma^{-1}(\g)&=1-\frac{1}{6}\g+\frac{1}{27}\g^2-0.0230696213\g^3
+0.0198868203\g^4 \cr &\quad -0.02245952\g^5+0.0303679\g^6,& \eqnn \cr
\eta(\g)&=\frac{8}{729}\g^2+0.0009142223\g^3+0.0017962229\g^4 \cr&\quad
-0.00065370 \g^5+0.0013878 \g^6,&\eqnn \cr}$$  
with the normalization:
$$\g =3g/(16\pi) .\eqnd\egNick $$
It has thus been possible to determine critical exponents quite accurately
(table 1).\par
To determine the equation of state or universal ratios of amplitudes
a new problem arises. In this framework it is
more difficult to calculate physical quantities in the ordered phase because
the theory is parametrized in terms of the disordered phase correlation length
$\xi=m^{-1}$ which is singular at $T_c$
(as well as correlation function normalization condition, eq.\egrzerom{}).
Let us consider the perturbative expansion of the scaling
equation of state \eVpotFeq~. For example at 
one-loop order for $d=3$ the function $F(z,g)$ is given by:
$$\eqalignno{F(z,g)& = z +\frac{1}{6}z^3 -\frac{1}{8\pi}gz\left[\left(1+z^2/2
\right)^{1/2}-1-z^2/4\right] &\eqnd\eoneloop\cr 
&= z +\frac{1}{6}z^3+\frac{1}{256\pi}gz^5-{1\over2^{13}\pi}gz^7+O(z^9),\cr} $$
where the subtractions, due to the mass and coupling normalizations,
are determined by the conditions \egrzerom{}.
This expression is adequate for the description of the disordered phase, but
all terms in the loopwise expansion become 
singular when $t$ goes to zero for fixed magnetization,
that corresponds to the limit $m\rightarrow 0$, $\varphi$ fixed
for the effective potential, or   
$z\to \infty$ for reduced variables (see eq.\ezscalvar).
In this regime we know  from eq.\ehscal~
(or from RG considerations)
that the equation of state is 
$$H(M,t=0)\propto M^{\delta}\ \Rightarrow\ F(z)\propto
z^\delta,\eqnd\ettozero$$ exactly.
\par
In the case of of the $\varepsilon$-expansion the very essence of the method 
is that one is doing perturbative expansions for  the theory {\it at}\/ the
critical point: it follows that the scaling relations (and thus the limit
eq.~\ettozero) are exactly satisfied order by order.
Moreover the change to the  variable $x\propto z^{-1/\beta}$ (more
appropriate for the regime $t\rightarrow 0$) gives an expression for 
$f(x)\propto F(x^{-\beta}) x^{\beta\delta}$ that is explicitly regular in
$x=0$ 
(Griffith's analyticity at the critical point): the singular powers of $\log
x$ induced by the change of variables cancel non trivially at each order,
leaving only regular corrections. 
\par
The situation changes when one deals with the perturbation theory at $d=3$
dimensions: for $g$ generic the system is no more at the critical point,
and consequently scaling properties are not satisfied order by order in $g$.
In particular the change to the Widom function $f(x)$ will introduce the
singular terms in $\log {x}$ that violates Griffith's analyticity.
An analogous problem arises if one first sums the series at $g=g^*$
before changing to the variable $x$.  
In this case the singular contributions (in the form of powers of $x$) do not
cancel, as a result of unavoidable numerical summation errors.
\par
Several approaches can be used to deal with the problem
of reaching the ordered phase. In \rBB~a method to calculate amplitude ratios
is proposed which has been also used in \refs{\rBBMN,\rMunster}.
\par
If we are concerned only with amplitude ratios another strategy is available
that in some sense bypasses the problem. Near the critical point physical
quantities have simple power law singularities in $t$. To reach the
coexistence curve, i.e.~$t<0, H=0$, it is possible to
proceed by analytic continuation in the complex $t$-plane. 
From equations \eqns{\ezvaria,\estaDfix} 
$$H(M,t)\propto t^{\beta\delta}F(z),\quad z\propto M t^{-\beta},$$
and the knowledge that at $M$ fixed $H(M,t)$ is regular at $T_c$ or $t=0$
we conclude that $t<0$ corresponds to $z$ complex
$$ t=\left\vert t\right\vert \e^{i\pi},\ \Rightarrow\  z=\left\vert
z\right\vert \e^{-i\pi\beta} .\eqnd\ezphase $$
The scaling variable $H(-t)^{-\beta\delta}$ is then given by: 
$$H(-t)^{-\beta\delta}=\e^{i\pi\beta\delta}F(z)=\left\vert F(z)\right\vert
.\eqnn $$
Finally the spontaneous magnetization is given in terms of the complex
zero $z_0$ of $F(z)$. \par
It is in particular possible to evaluate ratios of amplitudes of
singularities above and below $T_c$: we calculate the complex zero
$z_0(g)$ of $F(z,g)$ and substitute it in other 
quantities. The result is complex but its absolute value
 converges towards the
correct result. For example the ratio of amplitudes for the 
magnetic susceptibility (eq.~\estaDfix~) is given by: 
$$C^+ /C^- = \e^{-i\pi\gamma}F'\left(z_0\left(g
\right),g\right)/F'\left(0,g\right) =\left| F'\left(z_0\left(g
\right),g\right)\right| .\eqnn $$
We thus get a series expansion in $g$ for $C^+/C^-$ which can be summed
at $g=g^*$ with
techniques described in section \ssUQsum. 
However this method does not allow to calculate for $t$ small and thus is not
well suited to determine the full equation of state. Moreover if we want to
sum perturbation series with the method recalled in section \ssUQsum, we will
face the problem that the large order behaviour depends on the value of the
variable $z$ itself. 
\par
Encouraged by the results obtained 
within the $\varepsilon$-expansion scheme, we develop a different, more
powerful strategy, based on the parametric representation.
\medskip
{\it The parametric representation. Order dependent mapping (ODM).}
The problem that we face is the following: to reach the ordered region $t<0$
for the (summed) equation of state function $F(z,g^*)$, 
we must cross the point $z=\infty$ for which the exact behaviour is dictated
by eq.~\ettozero. The idea that the $\varepsilon$-expansion suggests 
is to introduce an new field variable $\theta$ and an auxiliary function
$h(\theta)$ defined as in \eqns{\ezmaptheta,\emagfparii}:  
in this way the exact function $h(\theta)$ will be regular near 
$\theta=1$ (i.e.~$z=\infty$) and up to the coexistence curve. However, the
approximate $h(\theta)$ that we obtain by summing perturbation theory at fixed
dimension, will still not be regular. The singular terms generated
by the the mapping eq.~\ezmaptheta~ at $\theta=1$ will not cancel exactly due
to summation errors. The last step we propose is to Taylor expand
the approximate expression of $h(\theta)$ around $\theta=0$ and to 
truncate the expansion, enforcing in this way regularity. The next question
then is to which order should we expand?
Since the coefficients of the $\theta$ expansion are in one to one
correspondence with the coefficients of small $z$ expansion of the function
$F(z,g^*)$ the maximal power of $\theta$ in $h(\theta )$, 
should be equal to the maximal power of $z$ which can be reasonably well
summed. Indeed although the small $z$ expansion of $F(z)$ at
each finite loop order in $g$ contains an infinite number of terms,
the determination of the coefficients of the higher powers of $z$ is
increasingly difficult. The reasons are twofold:\par
(i) The number of terms of the series in $g$ required to get an
accurate estimate of $F_l$ increases with $l$  (see section \ssUQsum).\par
(ii) At any finite order in $g$ the function $F(z)$ has spurious singularities
in the complex $z$ plane  (see e.g.~eq.~\eoneloop, $z^2=-2$) that dominate  
the behaviour of the coefficients $F_l$ for $l$ large.  \par
In view of these difficulties we have to ensure the fastest possible
convergence of the small $\theta$ expansion. For this purpose we use the
freedom in the choice of the arbitrary parameter $\rho$ in eq.\ezmaptheta:
we determine it to minimize the last term in the truncated small $\theta$
expansion, i.e.~increasing the importance of small powers of theta which are
more accurately determined. This is nothing but the application to this
particular example of the series summation method based on ODM \ref\rOMD{R.
Seznec and J. Zinn-Justin, {\it J. Math. Phys.} 20 (1979) 1398\semi 
J.C. Le Guillou and J. Zinn-Justin, {\it Ann. Phys. (NY)} 147 (1983) 57\semi
R. Guida, K. Konishi and H. Suzuki, {\it Ann. Phys. (NY)} 241 (1995) 152; 249
(1996) 109.}.
\par
This strategy applied to the available data, leads at leading
order for $h(\theta)$ to a polynomial of degree 5, whose coefficients are
given by the relations \eFHrelat{}:
$$h(\theta)=\theta[1+h_3(\rho)\theta^2+h_5(\rho)\theta^4]. \eqnd\ehtetv $$
For the range of admissible values for $F_5$ the coefficient $h_5$ of
$\theta^5$ given by eq.~\eFHrelat{b} has no real zero in $\rho$. It has a
minimum instead 
$$\rho^2=\rho_5^2={1\over12 F_5}(\gamma-2\beta) .\eqnd\erhomin$$
Substituting this value of $\rho$ into expression \ehtetv~we obtain the first
approximation for $h(\theta)$. At next order we look for a minimum $\rho_7$ of
$|h_7(\rho)|$. We find a polynomial either of degree 5 in $\theta$, when 
$h_7$ has a real zero, or of degree 7 when it has only a minimum. \par
We have not explored beyond $h_9(\rho)$ because already $F_9$ is too
poorly determined.
\par
Finally we want to point out that, to compute the amplitude ratios and thus 
to reach the ordered phase quantities we implement in the present
framework the idea of analytic continuation outlined in the previous
subsection. In particular the ordered phase at zero magnetic field will
correspond to the choice $\theta=\theta_0$ where $\theta_0$ will be the zero
of $h(\theta)$ closest to the origin.
\section The equation of state: series summation

Our first task is thus to determine the coefficients $F_{2l+1}$ as
accurately as possible. To sum the perturbation series the Borel--Leroy
\ref\rBorsom{The Borel summability of the $\phi^4$ in two and three dimensions
has 
been established in J.P. Eckmann, J. Magnen and R. S\'en\'eor, {\it Commun.
Math. Phys.} 39 (1975) 251\semi
J.S. Feldman and K. Osterwalder, {\it Ann. Phys. (NY)} 97 (1976) 80\semi
J. Magnen and R. S\'en\'eor, {\it Commun. Math. Phys.} 56 (1977) 237\semi
J.-P. Eckmann and H. Epstein,  {\it Commun. Math. Phys.} 68 (1979) 245.} 
transformation has been used, combined with a conformal mapping \ref\rLoef{
Summation of series by Borel transformation and mapping 
was proposed by J.J. Loeffel, {\it Saclay Report}, DPh-T/76/20 unpublished.} 
(a simplified version of method used in \rLGZJiii~for critical exponents).
Let ${\cal S}(g)$
be the function whose series has to be summed. We then transform the
series:\sslbl\ssUQsum 
$${\cal S}(g)=\sum_{k=0} {\cal S}_k g^k, \eqnd{\eseries} $$
into:
$$ {\cal S}(g)= \sum_{k=0}^{}B_{k}(b)
\int^{\infty}_{0}t^{b}\e^{-t}u^k(gt) \d t\,,\eqnd{\eBoreltr} $$
with:
$$ u(s)={ \sqrt{1+as}- 1\over \sqrt{1+as}+ 1} \,. \eqnn $$
The coefficients $B_k$ are calculated by expanding in powers of $g$ the
r.h.s.~of equation \eBoreltr~and identifying with expansion \eseries.
(For motivations and details see e.g.~\rbook). 
The constant $a$ (corresponding to the scaled variable 
$\g={3\over 16 \pi} g$ of \rNMB) has been determined by
the large order behaviour analysis \rbook,  
$$a(d=3) =0.147774232\ , \eqnd\eaLOB $$ 
for the perturbative expansion in $d=3$ dimensions.
The parameter $b$ is adjusted empirically to improve the convergence of the
transformed series: for example one looks for the intersection between the
results at two consecutive orders in $k$ (or the minimum of the difference),
or a point of least sensitivity. Moreover the value of $b$ has to stay in a
reasonable range around the value predicted by the large order behaviour.
It is also to be expected that the summation method will be efficient if the
coefficients ${\cal S}_k$ are already approaching the asymptotic large order
regime. This is what we are going to test first. The expectations are that 
for the coefficients $F_{2l+1}$ the situation will deteriorate with increasing
$l$: indeed the large order behaviour estimate has increasingly large
corrections. Moreover for $l$ large $F_{2l+1}$ becomes dominated by the
perturbative singularity of $F(z)$ at $z^2=-2$ while the summed function has
singularities at different locations.
\medskip
{\it Large order behaviour analysis \ref\rLGZJed{Many articles on this topic
are reprinted in {\it Large Order Behaviour of Perturbation Theory}, {\it
Current Physics} vol.~7, J.C. Le Guillou and J. Zinn-Justin eds.,
(North-Holland, Amsterdam 1990).}.} The series are taken from
\refs{\rBBMN,\rHaDo, \rRAJA}. The coefficients of the perturbative expansion
of 
$$F_l\equiv{1\over l!}g^{1-(l+1)/2}m^{(l+1)/2-3}\Gamma^{(l+1)}(p_i=0,m,g)=
 \sum_k F_{lk} g^k ,$$
have the large order behaviour:
$$\eqalign{
F_{lk} &\mathop{\sim}_{k\to\infty}{1\over l!} C_{l+1}\; 
(-{2\over 3 I_4})^{k-1+(l+1)/2}
\;\Gamma(k+l+3/2)\cr 
C_l&\equiv- {\e^{-9I_4 /(32 \pi)}\over \pi (2\pi)^{3/2}}
\left({4I_1^2 \over I_4 }\right)^{l/2} 
\left[{\bar D_1} \left({{3\over4} I_4\over I_6-I_4}\right)^3\right]^{-1/2}
.\cr }$$
The constant $I_4={9 a\over 32 \pi}$ ($a\equiv a(d=3)$ is given by
eq.~\eaLOB), is related to the action of the instanton which describes
the instability of the $\phi^4$ for negative coupling, while  $I_1=
31.691522$, $I_6=659.868352$,  
(related to integrals of power $1,6$ of the scaled instanton solution) 
and ${\bar D_1}=10.544$ (related to the functional determinant)
have been computed  in \ref\rPABRE{E. Br\'ezin and G. Parisi, {\it J. Stat.
Phys.} 19 (1978) 269.}.
%%%%%%%%%%%%%%%%%%TABLE BEGIN%%%%%%%%%%%%%%%%%%%%%%%%%%%%%%%%%%%%%%%
\midinsert
$$ \vbox{\elevenpoint\offinterlineskip\tabskip=0pt\halign to \hsize
{& \vrule#\tabskip=0em plus1em & \strut\ # \ 
& \vrule#& \strut #
& \vrule#& \strut #  
& \vrule#& \strut #  
& \vrule#& \strut #  
&\vrule#\tabskip=0pt\cr
\noalign{\centerline{Table 3} \tableskip}
\noalign{\centerline{\it Large Order Behaviour Analysis
: $\left. F_{lk} \right/ F_{lk}^{\rm as} $.}\tableskip}
\fileth
%
%title begin
%
height2.0pt& \omit&& \omit&& \omit&&\omit&& \omit&\cr
&$ \hfill l \hfill$&&$ \hfill k=2 \hfill$&&$ \hfill k=3 \hfill$&&$ \hfill k=4
\hfill$&&$ \hfill k=5 \hfill$&\cr  
%
%title end
%
height2.0pt& \omit&& \omit&& \omit&& \omit&& \omit&\cr
\fileth
%
%item begin
height2.0pt& \omit&& \omit&& \omit&& \omit&& \omit&\cr
&$ \hfill 5\hfill$&&$ \hfill  1.9646\hfill$&&$ \hfill 
 1.5894\hfill$&&$ \hfill 1.4226  \hfill$&&$\hfill 1.3498 \hfill$&\cr
%item end
%item begin
height2.0pt& \omit&& \omit&& \omit&& \omit&& \omit&\cr
&$ \hfill 7\hfill$&&$ \hfill  1.9680\hfill$&&$ \hfill 
 1.9965\hfill$&&$ \hfill 1.9603  \hfill$&&$\hfill 1.9204 \hfill$&\cr
%item end
%item begin
height2.0pt& \omit&& \omit&& \omit&& \omit&& \omit&\cr
&$ \hfill 9\hfill$&&$ \hfill  1.4443\hfill$&&$ \hfill 
 1.9879\hfill$&&$ \hfill 2.229  \hfill$&&$\hfill 2.333 \hfill$&\cr
%item end
%
%end
%
& \omit&& \omit&& \omit&& \omit&& \omit&\cr
\fileth}}$$
\endinsert
%%%%%%%%%%%%%%%%%%%%%%%%%%%TABLE END%%%%%%%%%%%%%%%%%%%%%%%%%%%%%%%%%%%%%%% 
Table 3 shows that, as anticipated, the asymptotic regime sets in later
when $l$ increases. We thus 
expect that the efficiency of the summation will correspondingly decrease, and
indeed this is what happens.
\medskip
{\it Summation.} Following an idea introduced in \rLGZJiii~for the summation
of the $\varepsilon$-expansion we have in addition made a
homographic transformation on the coupling constant $g$ to displace some 
singularities in the complex $g$-plane:
\def\param{p}
$$g=g'/(1+ \param g' /g^*). \eqnn $$
%We have looked for a value of the parameter $\param$ for which the results 
%where specially insensitive to the order $l$ and the parameter $b$. Then at
%$\param$ fixed  we have chosen for $b$ values for which the difference between
%two successive  approximants
%was minimal. 
We have looked for a value of the parameters $\param$ and $b$
for which the results were specially insensitive to the order $k$:
in practice we minimized the absolute differences of results corresponding to
three successive orders.
Finally to verify that our method gives results consistent with 
those of the previous analysis \rLGZJii~we have applied it to the RG
$\beta$-function and exponents. 
Figure 1 gives the last four orders for the
$\beta$-function as a function of the parameter $b$ and
at the minimal value $\param=0.196$.
One observes that the curves flatten with increasing order, as expected.
The central value slightly differs from the result given in \rLGZJii:
$$g^*=23.70\pm 0.05\ ,\quad{\rm with}\ \omega=\beta'(g^*)=0.79\pm0.01\ ,$$ 
and the apparent error is smaller. Note that a variation of $g^*$ of $0.05$
yields a variation of $0.04$ of the $\beta$-function.\par 
Figure 2 displays the results for the exponent $\gamma$ for the four last
orders, for $g^*=23.70$, and for  the minimal value $\param=.182$.
One obtains 
$$\gamma=1.2405\pm0.0012 \ .$$
With the same method one finds for the exponent $\beta$
$$\beta=0.3250\pm0.0015 \ .$$
One notices the prefect agreement with the result of table 1, although the
central value for $g^*$ has changed. The reason is that the central values 
given in \rLGZJii~for the exponents largely rely on the pseudo-$\varepsilon$
expansion, which avoids the explicit determination of $g^*$.
\midinsert
\epsfysize=7cm
\epsfxsize=12cm
\centerline{\epsfbox{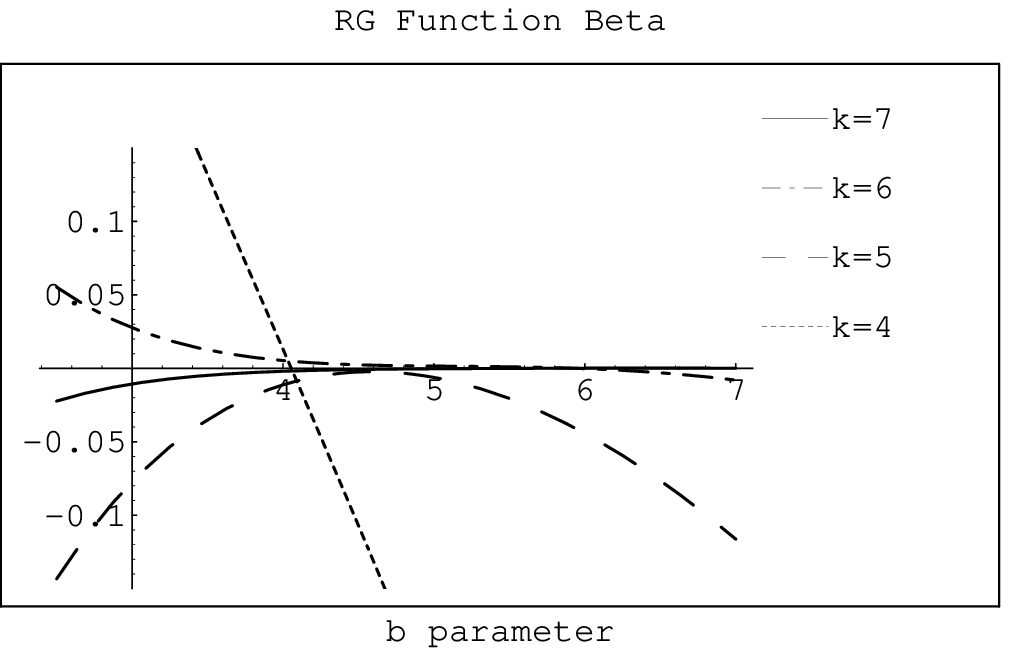}}
\figure{3.mm}{The RG $\beta$-function plotted vs.~the parameter $b$ for
successive orders $k$.} 
\endinsert
\midinsert
\epsfysize=7cm
\epsfxsize=12cm
\centerline{\epsfbox{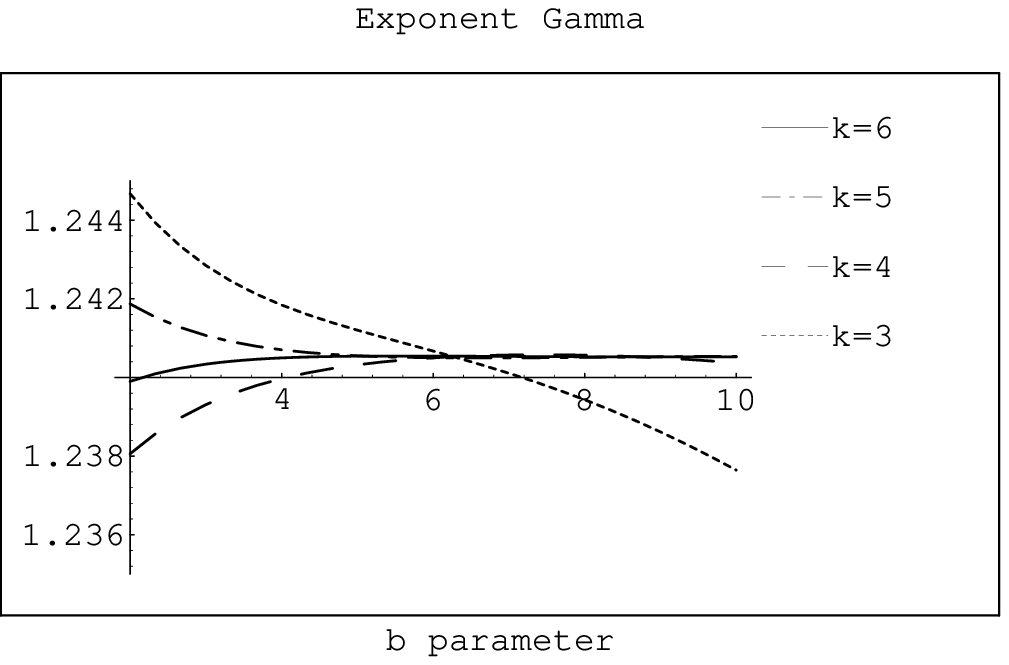}}
\figure{3.mm}{The exponent $\gamma$  plotted vs.~the parameter $b$ for
successive orders $k$.} 
\endinsert
\nref\rZLFish{
S. Zinn, S.-N. Lai, and M.E. Fisher, {\it Phys. Rev. E} 54 (1996) 1176.}
\section Numerical results: Equation of state

We first calculate the first coefficients of the small $z$ expansion of the
equation of state function $F(z)$
(or equivalently of the reduced effective potential, eq.~\eVpotFeq, summed
at the I.R. fixed point $g^*$).
% The same value of $\param$ which was optimal for the
%$\beta$-function and $\gamma$ happens to lead to good convergence 
%for $F_5$, as displayed in figure 3. 
In figure 3 we display the behaviour of $F_5$ in terms of $b$ at the minimal
value of the parameter $\param=.182$ (see section \ssUQsum~for definition of
parameters). 
\midinsert
\epsfysize=7cm
\epsfxsize=12cm
\centerline{\epsfbox{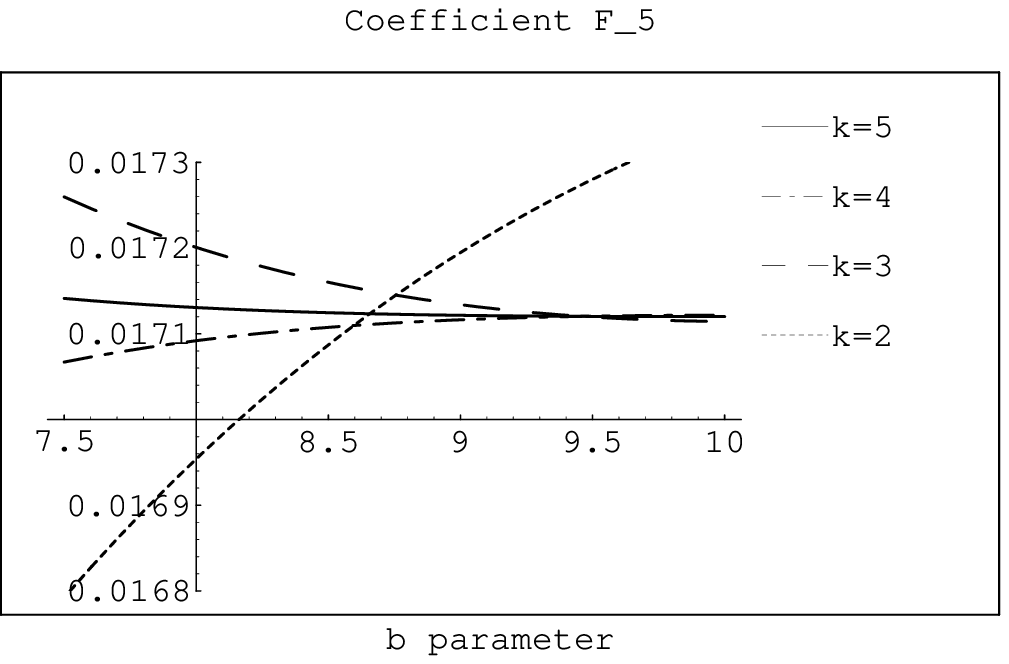}}
\figure{3.mm}{The coefficient $F_5$  plotted vs.~the parameter $b$ for
successive orders $k$.} 
\endinsert
Table 4 contains our results together with other published estimates of the
coefficients of the small $z$ expansion of $F(z)$.
For what concerns the $\varepsilon$-expansion, we have taken the
$\varepsilon$-expansion of $h(\theta)$, eqs.~\eqns{\estatparm,\edevbc}~ 
setting $\varepsilon=1$. We have then calculated the coefficients $F_{2l+1}$,
using the numerical values of the more accurately determined
exponents  $\gamma,\beta$ (second line of table 1). This procedure has led to
much stabler results for $F_{2l+1}$ than a direct summation of the
$\varepsilon$-expansion. The quoted errors are nevertheless
only indicative (and subjective) because $h(\theta)$ is known only to order
$\varepsilon^2$ for $h_3$ and $\varepsilon^3$ for $h_5$.
Note that although $g^*$ is known only to order
$\varepsilon^2$ \ref\BLGZJ{E. Br\'ezin, J.C. Le Guillou and J. Zinn-Justin,
{\it Phys. Rev.} D8 (1973) 434.}
$$g^*(\varepsilon)=\frac{1}{3}\Gamma(d/2)(4\pi)^{d/2}\varepsilon
\left(1+\frac{61}{54}\varepsilon\right)+O(\varepsilon^3),$$
evaluation of the expression at $\varepsilon=1$
gives nevertheless a number of the right order of magnitude $g^*=28.$
\nref\rsokolov{A.I. Sokolov, V.A. Ul'kov and E.V. Orlov, presented at the
conference {\it Renormalization Group 96}, Dubna 1996,
data obtained from Pad\'e Borel summation up to order $g^3$, private
communication. See also A.I. Sokolov,{\it Fizika Tverdogo Tela (Solid State
Physics)} 38(1996) 640.}
\nref\rreisz{T. Reisz,{\it Phys. Lett.} B360 (1995) 77.}
\nref\rbuco{Preliminary results from High Temperature Series
for free energy at order $O(\beta^17)$,  P. Butera and M. Comi
in preparation.}
\nref\rtsypin{M.M. Tsypin,{\it Phys. Rev. Lett.} 73 (1994) 2015.}
\nref\rkimlandau{J-K Kim and D.P. Landau, {\bf hep-lat/9608072}.}
\nref\rwetterich{N. Tetradis and C. Wetterich, {\it Nucl. Phys.} 
B422 (1994) 541.}
\nref\rmorris {T. Morris, data obtained from order derivative squared
approximation to the Exact Renormalization Group,
private communication.}
\nref\rfishertarkoa{M.E. Fisher and H.B. Tarko, {\it Phys. Rev.} B11 (1975)
1131.} 
\nref\rgauntdomb{D.S. Gaunt and C. Domb, {\it J. Phys.C} 3 (1970) 1442.}
\nref\rwetterichb{J. Berges, N. Tetradis and C. Wetterich,
{\it Phys. Rev. Lett.} 77 (1996) 873.}
\midinsert
$$ \vbox{\elevenpoint\offinterlineskip\tabskip=0pt\halign to \hsize
{& \vrule#\tabskip=0em plus1em & \strut\ # \ 
& \vrule#& \strut #
& \vrule#& \strut #  
& \vrule#& \strut #  
& \vrule#& \strut #  
&\vrule#\tabskip=0pt\cr
\noalign{\centerline{Table 4} \tableskip}
\noalign{\centerline{\it Equation of state.}\tableskip}
\fileth
height2.0pt& \omit&& \omit&& \omit&&\omit&& \omit&\cr
&$ \hfill  \hfill$&&$ \hfill g^* \hfill$&&$ \hfill
F_5  \hfill$&&$ \hfill F_7
\hfill$&&$ \hfill F_9 \hfill$&\cr 
height2.0pt& \omit&& \omit&& \omit&& \omit&& \omit&\cr
\fileth
height2.0pt& \omit&& \omit&& \omit&& \omit&& \omit&\cr
&$ \hfill \varepsilon{-\rm exp.} \hfill$&&$ \hfill 28.
\hfill$&&$ \hfill 
0.0176\pm 0.0004 \hfill$&&$ \hfill (4.5\pm0.3)\times 10^{-4}  \hfill$&&$
\hfill -(3.2\pm0.2) \times 10^{-5} \hfill$&\cr
height2.0pt& \omit&& \omit&& \omit&& \omit&& \omit&\cr
&$ \hfill d=3 \hfill$&&$ \hfill 23.70\pm 0.05  \hfill$&&$ \hfill
0.01712\pm 0.00006 \hfill$&&$ \hfill (4.96\pm 0.49)\times 10^{-4}   \hfill$&&$
\hfill -(6.2\pm4.0)\times 10^{-5} \hfill$&\cr 
height2.0pt& \omit&& \omit&& \omit&& \omit&&\omit&\cr
&$ \hfill   d=3 ~\rsokolov \hfill$&&$ \hfill 23.73 \hfill$&&$ \hfill
.01727
\hfill$&&$ \hfill .0010 \hfill$&&$ \hfill
  \hfill$&\cr
height2.0pt& \omit&& \omit&& \omit&& \omit&& \omit&\cr
&\hfill   HT~\rreisz \hfill&&$ \hfill 23.72 \pm 1.49 \hfill$&&$ \hfill
0.0205\pm 0.0052 
\hfill$&&$ \hfill   \hfill$&&$ \hfill \hfill$&\cr
height2.0pt& \omit&& \omit&& \omit&& \omit&& \omit&\cr
&\hfill  HT~\rZLFish \hfill&&$ \hfill  24.45\pm0.15\hfill$&&$ \hfill
.017974\pm.00015
\hfill$&&$ \hfill  \hfill$&&$ \hfill
  \hfill$&\cr
height2.0pt& \omit&& \omit&& \omit&& \omit&& \omit&\cr
&\hfill  HT~\rbuco \hfill&&$ \hfill 23.69 \hfill$&&$ \hfill
.0166
\hfill$&&$ \hfill .00055 \hfill$&&$ \hfill
.00001  \hfill$&\cr
height2.0pt& \omit&& \omit&& \omit&& \omit&& \omit&\cr
&\hfill   MC~\rtsypin \hfill&&$ \hfill 23.3 \pm 0.5  \hfill$&&$ \hfill
0.0227\pm0.0026 
\hfill$&&$ \hfill  \hfill$&&$ \hfill  \hfill$&\cr
height2.0pt& \omit&& \omit&& \omit&& \omit&& \omit&\cr
& \hfill MC~\rkimlandau \hfill&&$ \hfill 24.5\pm.2 \hfill$&&$
\hfill 0.027\pm0.002 
\hfill$&&$ \hfill 0.00236\pm.00040  \hfill$&&$ \hfill  \hfill$&\cr
height2.0pt& \omit&& \omit&& \omit&& \omit&& \omit&\cr
& \hfill   ERG~\rwetterich \hfill&&$ \hfill 28.9  \hfill$&&$ \hfill
0.016 
\hfill$&&$ \hfill 4.3\times 10^{-4} \hfill$&&$ \hfill  \hfill$&\cr
height2.0pt& \omit&& \omit&& \omit&& \omit&& \omit&\cr
&\hfill ERG~\rmorris \hfill&&$ \hfill 20.7\pm0.2  \hfill$&&$ \hfill
0.0173  \pm 0.0001 
\hfill$&&$ \hfill (5.0\pm 0.2)\times 10^{-4} \hfill$&&$ \hfill
-(4.\pm 2.)\times 10^{-5}  \hfill$&\cr
height2.0pt& \omit&& \omit&& \omit&& \omit&& \omit&\cr
\fileth}}$$
\endinsert
\medskip
{\it Parametric representation.} We then determine by the ODM method the
coefficient $\rho$ and the function $h(\theta)$, as explained in section
\ssUQfixD. We obtain successive approximations in the form of polynomials of
increasing degree for $h(\theta)$. Note that we here have a simple test of the
relevance of the ODM method. Indeed, once $h(\theta)$ is determined,
assuming the values of the critical exponents $\gamma$ and $\beta$, we can
recover a function $F(z)$ which has an expansion to all orders in $z$. As a
result we obtain a prediction for the coefficients $F_{2l+1}$ which have not
yet been taken into account to determine $h(\theta)$. The relative difference
between the predicted value and the calculated one gives an idea about the
accuracy of the ODM method. Indeed from the values $F_5=0.01712,
\gamma=1.2405, \beta=0.3250$, we obtain
$$F_7=4.94\times 10^{-4} ,\quad F_9=-3.2\times 10^{-5},\quad F_{11}=8.5\times 
10^{-8}\ \cdots \ . $$
We note that the value for $F_7$ is very close to the central value
we find by direct series summation, while the value for $F_9$ is within the
error bars. This result gives us confidence in our method. It also shows that
the value of $F_9$ obtained by direct summation contains very little new
information, it provides only a consistency check.
Therefore the simplest representation of the equation of state, consistent
with all data,  is given by
$$h(\theta)/\theta=1-0.76147(36)\;\theta^2+7.74(11)\times 10^{-3}\;\theta^4,
\eqnd\mainresult$$
(errors on the last digits in parentheses) that is obtained 
from $\rho^2=2.8743\ $ (fixed according eq.~\erhomin).
This expression of $h(\theta)$ has a zero at 
$$\theta_0^2=1.331, \eqnn$$
to which corresponds the value of the complex root $z_0$ of $F(z)$,
$|z_0|=2.801$ (the phase, given by eq.~\ezphase, is $-i\pi \beta$).
The coefficient of $\theta^7$ in eq.~\mainresult~is smaller than $10^{-3}$.
Note that for the largest value of $\theta^2$ which corresponds to 
$\theta_0^2$, the $\theta^4$ term is still a small correction. 
Finally note that the corresponding values for the $\varepsilon$-expansion
are $h_3=-0.72$, $h_5=0.013$. These values are reasonably 
consistent between them, since a small change in the value of $\rho$ with a
correlated change in $h_3$ induces a very small change in physical quantities.
\medskip
The Widom scaling function $f(x)$, eq.~\ehscal, 
(with $f(-1)=0$ and $f(0)=1$) can easily be derived by (numerically) solving
the following system:
$$
\left\lbrace\eqalign{
&f(x)=\theta^{-\delta} h(\theta)/h(1)\cr 
&x=  \left({1-\theta^2\over 1-\theta_0^2 }\right) 
\left( {\theta_0\over \theta} \right)^{1/\beta} 
}\right. \eqnd\eWidomnum 
$$
Our results are displayed in fig.~4, where they are compared with
$\varepsilon^2$ (Linear Parametric Model) and $\varepsilon^3$ predictions,
High Temperature series ("quintic" fit amplitude ratios results of
\rfishertarkoa~and Pad\'e approximants of bcc lattice results of \rgauntdomb) 
as well with Exact Renormalization Group (a fit reported 
in \rwetterichb~that approximates numerical data up to $1\% $). Note that the
rapid increase of relative errors for $x<0$ is not very significative since
the absolute values are small.
\midinsert
\epsfysize=7cm
\epsfxsize=12cm
\centerline{\epsfbox{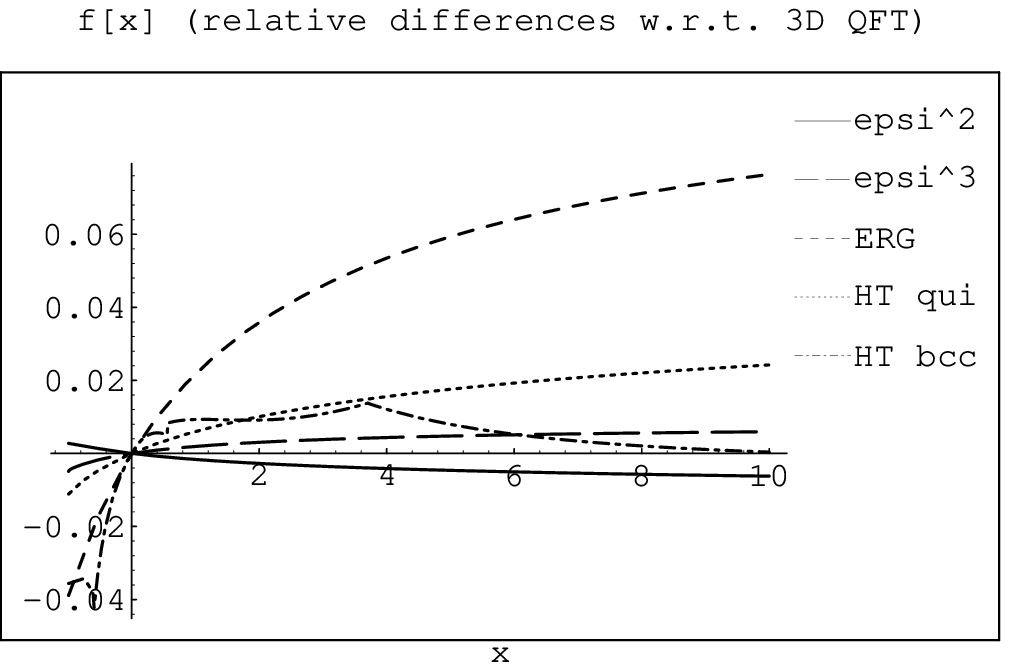}}
\figure{3.mm}{Widom scaling function $f(x)$: 
relative errors with respect 3D QFT plotted vs.~$x$.}
\endinsert
One notices the general agreement between different predictions 
within $6\% $ up to $x=10$. 
In particular the general agreement between our predictions
and the $\varepsilon$-expansion is striking.
The main disagreement with other predictions comes from the region 
$x\rightarrow \infty$, i.e.~from the small magnetization region, where 
our predictions should be specially reliable. 
Finally let us notice that the results \refs{\rfishertarkoa,\rgauntdomb} are
based some obsolete estimates of critical amplitudes
($\gamma=5/4,\;\beta=5/16$).   
\nref\rliu{A.J. Liu and M.E. Fisher, {\it Physica} A156 (1989) 35.}
\nref\raharony{A. Aharony and P.C. Hohenberg, {\it Phys. Rev.} B13 (1976)
3081; {\it Physica} 86-88B (1977) 611.}
\nref\rprivman{V.Privman, P.C. Hohenberg, A. Aharony, {\it Universal Critical
Point Amplitude Relations}, in Phase Transitions and Critical Phenomena
vol.~14, C.~Domb and J.L.~Lebowitz eds., (Academic Press 1991).}
\section Amplitude ratios. Conclusion

We finally use the expressions \eqns{\eratAA{--}\eratRchi} with the critical
exponents determined by earlier work (from longer series) to calculate various
amplitude ratios. As a comparison we apply the same procedure to the
$\varepsilon$-expansion at order $\varepsilon^3$ (using the values of
exponents determined from the $\varepsilon$-expansion at order
$\varepsilon^5$). It is necessary to again stress that as in table 4 the
errors we then quote are largely tentative since the order $\varepsilon^3$ is
the first order where some reliable results may be expected. \par
Table 5 contains a comparison of amplitude ratios as obtained from
RG, lattice calculations and experiments on binary
mixtures, liquid--vapour, uniaxial magnetic systems.\sslbl\ssUQampl
\midinsert
$$ \vbox{\elevenpoint\offinterlineskip\tabskip=0pt\halign to \hsize
{& \vrule#\tabskip=0em plus1em & \strut\ # \ 
& \vrule#& \strut #
& \vrule#& \strut #  
& \vrule#& \strut #  
& \vrule#& \strut #  
&\vrule#\tabskip=0pt\cr
\noalign{\centerline{Table 5} \tableskip}
\noalign{ \centerline {\it Amplitude ratios.}
\tableskip} 
\fileth
height2.0pt& \omit&& \omit&& \omit&&\omit&& \omit&\cr
&$ \hfill $&&$ \hfill A^+/ A^-
\hfill$&&$ \hfill C^+/C^- \hfill$&&$ \hfill
R_c \hfill$&&$ \hfill R_\chi\hfill$&\cr
height2.0pt& \omit&& \omit&& \omit&& \omit&& \omit&\cr
\fileth
height2.0pt& \omit&& \omit&& \omit&& \omit&& \omit&\cr
%%%%%
&$ \hfill \varepsilon-{\rm exp.}~(a) \hfill$&&$ \hfill 0.524\pm 0.010
\hfill$&&$ \hfill 
4.9 \hfill$&&$ \hfill  \hfill$&&$
\hfill 1.67 \hfill$&\cr
%%%%%
%%%%%
&$ \hfill \varepsilon-{\rm exp.}~(b) \hfill$&&$ \hfill 0.547\pm 0.021
\hfill$&&$ \hfill 
4.70\pm 0.10 \hfill$&&$ \hfill 0.0585\pm0.0020 \hfill$&&$
\hfill 1.649\pm0.021 \hfill$&\cr
%%%%%
height2.0pt& \omit&& \omit&& \omit&&
\omit&& \omit&\cr &$ \hfill d=3~{\rm fixed} ~(a)\hfill$&&$ \hfill 0.541\pm
0.014 \hfill$&&$ \hfill  
4.77\pm 0.30\hfill$&&$ \hfill 0.0594\pm 0.001\hfill$&&$ \hfill 1.7
\hfill$&\cr 
height2.0pt& \omit&& \omit&& \omit&&
\omit&& \omit&\cr &$ \hfill d=3~{\rm fixed} ~(b) \hfill$&&$ \hfill 0.536\pm
0.019 \hfill$&&$ \hfill  
4.82\pm 0.10\hfill$&&$ \hfill 0.0576\pm 0.0020\hfill$&&$ \hfill 1.674\pm0.019
\hfill$&\cr 
height2.0pt& \omit&& \omit&& \omit&& \omit&&\omit&\cr
&$ \hfill \hbox{HT series} \hfill$&&$ \hfill 0.523\pm 0.009 \hfill$&&$ \hfill
4.95 \pm 0.15 
\hfill$&&$ \hfill 0.0581\pm0.0010 \hfill$&&$ \hfill 1.75
\hfill$&\cr 
height2.0pt& \omit&& \omit&& \omit&& \omit&& \omit&\cr
&$ \hfill {\rm bin.~mix.} \hfill$&&$ \hfill 0.56\pm0.02
\hfill$&&$ \hfill 
4.3\pm 0.3 \hfill$&&$ \hfill 0.050\pm0.015  \hfill$&&$ \hfill  1.75\pm0.30
\hfill$&\cr  
height2.0pt& \omit&& \omit&& \omit&& \omit&& \omit&\cr
&$ \hfill {\rm liqu.-vap.} \hfill$&& \hfill 0.48--0.53 \hfill&& \hfill 
4.8{--}5.2 \hfill&&$ \hfill 0.047\pm0.010  \hfill$&&$ \hfill  1.69\pm0.14
\hfill$&\cr 
height2.0pt& \omit&& \omit&& \omit&& \omit&& \omit&\cr
&$ \hfill {\rm magn.~syst.} \hfill$&& \hfill 0.49{--}0.54 \hfill&&$ \hfill 
4.9\pm 0.5 \hfill$&&$ \hfill   \hfill$&&$ \hfill  
\hfill$&\cr 
height2.0pt& \omit&& \omit&& \omit&& \omit&& \omit&\cr
\fileth}}$$
\endinsert
Results for $\varepsilon$-expansion ({\it a}) are taken from
\rbervil~and \rNA~(direct Pad\'e summation of each corresponding series),
while ({\it b}) are our results, obtained by first summing
$h(\theta)$ and then computing ratios by use of
eqs.~\eqns{\eratCC{--}\eratRchi}. 
The results $d=3$ fixed dimension ({\it a}) are taken from 
\rBBMN~and refers to direct summation up to $O(g^5)$ while $d=3$ ({\it b})
is the present work. High Temperature results are taken  from \rliu~($R_\chi$
from \raharony). Experimental data are  extracted from \rprivman, to which we
refer for an updated and wider list of results (and references). 
We note the general consistency of the results obtained by
different methods.
%%new%%%%%%
In addition in table 7 we compare our predictions with the results of 
\rZLFish~concerning the following the universal ratios: 
$$R_0={(C^+_4)^2\over C^+ C_6^+}, \;\;\; R_3=-{C^-_3 B\over (C^-)^2},
\;\;\;\; {C_4^+ \over C_4^-}$$
where the $C^\pm_k$ are defined by 
${\partial^k \chi /\partial H^{k}} =C_{k+2}^{\pm}
|t|^{-\gamma-k\beta\delta}$ and the other quantities have been defined in  
section \ssefact. After some algebra one finds the relation 
$$R_0={1\over 10} (1-12 F_5)^{-1}$$
that has thus been used in Table 4. 
%%%%%%%%%%%%%%%%%%TABLE BEGIN%%%%%%%%%%%%%%%%%%%%%%%%%%%%%%%%%%%%%%%
\midinsert
$$ \vbox{\elevenpoint\offinterlineskip\tabskip=0pt\halign to \hsize
{& \vrule#\tabskip=0em plus1em & \strut\ # \ 
& \vrule#& \strut #
& \vrule#& \strut #  
& \vrule#& \strut #  
&\vrule#\tabskip=0pt\cr
\noalign{\centerline{Table 6} \tableskip}
\noalign{\centerline{\it Other amplitude ratios.}\tableskip}
\fileth
%
%title begin
%
height2.0pt& \omit&& \omit&& \omit&&\omit &\cr
&\omit && $\hfill R_0 \hfill$&&$ \hfill R_3 \hfill$&&$ \hfill {C_4^+/
C_4^-}\hfill$&\cr  
%
%title end
%
height2.0pt& \omit  && \omit&& \omit&& \omit&\cr
\fileth
%
%item begin
height2.0pt& \omit&& \omit&& \omit&& \omit&\cr
&$ \hfill {\rm HT\ series}~\rZLFish\hfill$&&$ \hfill  0.1275\pm0.0003\hfill$&&
$ \hfill 6.4\pm0.2 \hfill$&&$ \hfill -9.0\pm 0.3 \hfill$&\cr
%item end
%item begin
height2.0pt& \omit&& \omit&& \omit&& \omit&\cr
& \hfill d=3 fixed \hfill&&$ \hfill  0.12586\pm0.00012\hfill$&&
$ \hfill 6.10\pm0.06 \hfill$&&$ \hfill -9.2\pm 0.6 \hfill$&\cr
%item end
& \omit&& \omit&& \omit&&  \omit&\cr
& \hfill $\varepsilon$-expansion \hfill&&$ \hfill  0.1268\pm0.0008\hfill$&&
$ \hfill 6.11\pm0.10 \hfill$&&$ \hfill -8.3\pm 1.0 \hfill$&\cr
%item end
& \omit&& \omit&& \omit&&  \omit&\cr
\fileth}}$$
\endinsert
%%%%%%%%%%%%%%%%%%%%%%%%%%%TABLE END%%%%%%%%%%%%%%%%%%%%%%%%%%%%%%%%%%%%%%%% 

\medskip
{\it Conclusion.} Working within the framework of renormalized quantum field
theory and renormalization group, we have shown that the presently available
five loop series allow,  after proper summation, to determine with reasonable
accuracy the complete scaling equation of state for 3D Ising-like systems.
The parametric representation of the equation of state plays a central role in
our analysis. As a consequence new estimates of some amplitude
ratios have been obtained. Clearly a similar strategy could be applied to
other quantities in a magnetic field, in the scaling region. 
We want 
to  stress that an extension of the $\varepsilon$-expansion of the equation of
state for $N=1$ to order $\varepsilon^4$ or even better $\varepsilon^5$,
that does not seem an impossible task,
would significantly improve the $\varepsilon$-expansion estimates and
would therefore be extremely useful.
Moreover the present approach could be extended to systems in the
universality class of the $(\phi^2)^2_3$ field theory for higher $N$, provided
expansions of the renormalized effective potential at high enough order are
computed.   
\medskip
{\bf Acknowledgments.} The authors are indebted to V.~Dohm for
signaling them the misprints in \rBBMN, to B.~Nickel for sending them a data
file and to C.~Bervillier for careful reading of the manuscript. 
One of the author (R.~G.) wants also to thank A.K.~Rajantie and F.~Bernardeau
for useful information, and the INFN group of Genova for its kind
hospitality. The work of R.~G.~is supported by an EC TMR grant, 
contract N$^o$ ERB-FMBI-CT-95.0130. 
%%%%%%%%%%%RIC Bibliography%%%%%%%%%%%%%%%%%%%%%%%%%%%%%%%%%%%%%%%%%%%%
%

\vfill
\eject
\def\appendixname{Perturbative expansion of the effective potential}

\appendix{}

%%%%%%%%%%%%%%%%%%%RIC DEFINITIONS%%%%%%%%%%%%%%%%%%%%%%%%%
\def\ls{{\lambda_4}}
\def\ms{{m_{s}}}
\def\l{{l}}
\def\M{{M}}
\def\MS{{\overline{\rm MS}}}
%%%%%%%%%%%%%%%%%%%%%%%%%%%%%%%%%%%%%%%%%%%%%%%%%%%%%%%%%%%%

\section{\appendixname}

%$$\eqalign{
%f_3&=\frac{1}{34992(3 + x)}\left[L^3(2187 + 810x + 27x^2 )\right. \cr
%&\quad  + L^2\left( 12150 + 4428x + 450x^2 \right) \cr
%&\quad + L\bigl(9702 + x\left( 7440 - 7776\zeta(3) \right) \cr 
%& \quad + x^2\left( 1402 - 2592\zeta(3) \right)  \bigr)  \cr
%&\quad\left. + x^2\left( -324 - 648\lambda  + 1296\zeta(3) \right) \right]\cr
%} $$
%with
%$$L=\log (3+x).$$ 
%%%%%%%%%%%%%%%%%%coefficient tableckup 20/9/96%%%%%%%
%%%%%%%%%%%%%%%%%%TABLE BEGIN%%%%%%%%%%%%%%%%%%%%%%%%%%%%%%%%%%%%%%%
\midinsert
$$ \vbox{\elevenpoint\offinterlineskip\tabskip=0pt\halign to \hsize
{& \vrule#\tabskip=0em plus1em & \strut\ # \ 
& \vrule#& \strut #
& \vrule#& \strut #  
& \vrule#& \strut #  
& \vrule#& \strut #  
&\vrule#\tabskip=0pt\cr
\noalign{\centerline{Table 7} \tableskip}
\noalign{\centerline{\it Coefficients $F_{blk}$.}\tableskip}
\fileth
%
%title begin
%
height2.0pt& \omit&& \omit&& \omit&&\omit&& \omit&\cr
&$ \hfill b \hfill$&&$ \hfill l \hfill$&&$ \hfill k=0  \hfill$&&$ \hfill k=1 \hfill$&&$ \hfill k=2 \hfill$&\cr 
%
%title end
%
height2.0pt& \omit&& \omit&& \omit&& \omit&& \omit&\cr
\fileth
%
%item begin
%height2.0pt& \omit&& \omit&& \omit&& \omit&& \omit&\cr
%&$ \hfill B \hfill$&&$ \hfill  L\hfill$&&$ \hfill 
% K0\hfill$&&$ \hfill K1  \hfill$&&$\hfill K2 \hfill$&\cr
%item end
%item begin
height2.0pt& \omit&& \omit&& \omit&& \omit&& \omit&\cr
&$ \hfill 1\hfill$&&$ \hfill  0\hfill$&&$ \hfill 
 -{1/(12 \pi) }\hfill$&&$ \hfill   \hfill$&&$\hfill  \hfill$&\cr
%item end
%item begin
height2.0pt& \omit&& \omit&& \omit&& \omit&& \omit&\cr
&$ \hfill 2 \hfill$&&$ \hfill  0\hfill$&&$ \hfill{1/( 128 \pi^2)} 
 \hfill$&&$ \hfill   \hfill$&&$\hfill  \hfill$&\cr
%item end
%item begin
height2.0pt& \omit&& \omit&& \omit&& \omit&& \omit&\cr
&$ \hfill 2 \hfill$&&$ \hfill  1\hfill$&&$ \hfill 
 \hfill$&&$ \hfill -{1/( 96\pi^2)}   \hfill$&&$\hfill 
 \hfill$&\cr
%item end
%item begin
height2.0pt& \omit&& \omit&& \omit&& \omit&& \omit&\cr
&$ \hfill 3 \hfill$&&$ \hfill  0\hfill$&&$ \hfill 
 (5+8\log{{3\over4}})/(6144\pi^3)
\hfill$&&$ \hfill {1/( 768\pi^3)} \hfill$&&$\hfill  \hfill$&\cr
%item end
%item begin
height2.0pt& \omit&& \omit&& \omit&& \omit&& \omit&\cr
&$ \hfill 3 \hfill$&&$ \hfill  1\hfill$&&$ \hfill 
 {k_3/( 6912\pi^3)}\hfill$&&$ \hfill   \hfill$&&$\hfill  \hfill$&\cr
%item end
%item begin
height2.0pt& \omit&& \omit&& \omit&& \omit&& \omit&\cr
&$ \hfill 3 \hfill$&&$ \hfill  2\hfill$&&$ \hfill 
{ k_4/( 20736\pi^3)}\hfill$&&$ \hfill   \hfill$&&$\hfill  \hfill$&\cr
%item end
%item begin
height2.0pt& \omit&& \omit&& \omit&& \omit&& \omit&\cr
&$ \hfill 4 \hfill$&&$ \hfill  0\hfill$&&$ \hfill 
 0.449205291\times10^{-6}\hfill$&&$ \hfill 0.122593698\times10^{-5} \hfill$&&$\hfill
 \hfill$&\cr
%item end
%item begin
height2.0pt& \omit&& \omit&& \omit&& \omit&& \omit&\cr
&$ \hfill 4 \hfill$&&$ \hfill  1\hfill$&&$ \hfill 
 -.115567060\times10^{-5}\hfill$&&$ \hfill -{1/( 18432\pi^4)}  \hfill$&&$\hfill 
\hfill$&\cr
%item end
%item begin
height2.0pt& \omit&& \omit&& \omit&& \omit&& \omit&\cr
&$ \hfill 4 \hfill$&&$ \hfill  2\hfill$&&$ \hfill 
0.100772569\times10^{-5} \hfill$&&$ \hfill   \hfill$&&$\hfill  \hfill$&\cr
%item end
%item begin
height2.0pt& \omit&& \omit&& \omit&& \omit&& \omit&\cr
&$ \hfill 4 \hfill$&&$ \hfill  3\hfill$&&$ \hfill 
 -.300586776\times10^{-6}\hfill$&&$ \hfill   \hfill$&&$\hfill  \hfill$&\cr
%item end
%item begin
height2.0pt& \omit&& \omit&& \omit&& \omit&& \omit&\cr
&$ \hfill 5 \hfill$&&$ \hfill  0\hfill$&&$ \hfill 
 -.451638912\times10^{-7}\hfill$&&$ \hfill  
({3-8\log{{3\over4}})/( 1179648\pi^5)}
 \hfill$&&$\hfill -{1/( 294912\pi^5)}
 \hfill$&\cr
%item end
%item begin
height2.0pt& \omit&& \omit&& \omit&& \omit&& \omit&\cr
&$ \hfill 5 \hfill$&&$ \hfill  1\hfill$&&$ \hfill 
 0.806878091\times10^{-7}\hfill$&&$ \hfill {k_3/(1327104\pi^5)}  
\hfill$&&$\hfill  \hfill$&\cr
%item end
%item begin
height2.0pt& \omit&& \omit&& \omit&& \omit&& \omit&\cr
&$ \hfill 5 \hfill$&&$ \hfill  2\hfill$&&$ \hfill 
 -.100671980\times10^{-6}\hfill$&&$ \hfill {k_4/(1327104\pi^5)}  
\hfill$&&$\hfill  \hfill$&\cr
%item end
%item begin
height2.0pt& \omit&& \omit&& \omit&& \omit&& \omit&\cr
&$ \hfill 5 \hfill$&&$ \hfill  3\hfill$&&$ \hfill 
 0.647972392\times10^{-7}\hfill$&&$ \hfill   
\hfill$&&$\hfill  \hfill$&\cr
%item end
%item begin
height2.0pt& \omit&& \omit&& \omit&& \omit&& \omit&\cr
&$ \hfill 5 \hfill$&&$ \hfill  4\hfill$&&$ \hfill 
 -.165599191\times10^{-7}\hfill$&&$ \hfill   
\hfill$&&$\hfill  \hfill$&\cr
%item end

%
%end
%
& \omit&& \omit&& \omit&& \omit&& \omit&\cr
\fileth}}$$
\endinsert
%%%%%%%%%%%%%%%%%%%%%%%%%%%TABLE END%%%%%%%%%%%%%%%%%%%%%%%%%%%%%%%%%%%%%%%% 
Our starting point is the  renormalized five loop effective potential in
Minimal Subtraction ($\MS$) scheme, that  can  be written in the form:
$$
\eqalignno{
&V_{\MS}(\phi,\ms,\ls,\bar\mu)\equiv{\ms^2 \over 2} \phi^2+{\ls\over
4!}\phi^4\cr 
&\quad +\M^3\sum_{b=1}^{5} \sum_{l=1}^{b-1}\sum_{k=0}^{2}F_{blk}
\left({\ls\over\M}\right)^{b-1} \left({ {\ls\over 2} \phi^2\over M^2 
}\right)^{\l} \left(\log{{k_1 \bar\mu\over M}}\right)^k&\eqnd\msveff\cr 
}
$$
where $\ms$ ($\ls$) are the renormalized mass (coupling), 
and we defined $\M^2\equiv\ms^2+{\ls\over 2} \phi^2$, $k_1\equiv \sqrt{\e}/3
$. The constant $F_{blk}$ are reported in Table 7.

Analytical results up to three loops have been obtained  by \rRAJA, and
can be  parametrized by the two constants
$$\eqalign{
&k_3\equiv-{9\over2}+{9\pi^2\over4}-{27\over2}\bigl(\log(4/3)\bigr)^2-27{\rm
Li}_2(1/4)\cr 
&= 9.36271548728022813982\cdots\cr
\cr
&k_4\equiv-{27\pi^2\over8}+{81\over4}\bigl(\log(4/3)\bigr)^2+54 \log(4/3)
+{81\over2}{\rm Li}_2(1/4)\cr
&-{54\over \sqrt{2}} \int_0^1{\d x\over\sqrt{3-x^2}}
\left(\log{3\over 4} +\log{{3+x\over 2+x}}-{x^2\over 4-x^2}  \log{{4\over
2+x}}  +{x\over 2+x}\log{{3+x\over 3}}  \right)\cr
&=-6.43307044049269064141\cdots
\cr} $$
These expressions involve the dilogarithm or Spence function
$${\rm Li}_2(z)=\int_z^0{\log(1-t)\over t}\d t\ .$$
We have obtained the analytic expressions of some four and five loop
coefficients  by
imposing the $\MS$~RG equation on the effective potential,
 that in our notations reads:
$$\left(\bar\mu{\del\over \del \bar\mu}+{1\over 12(4\pi)^2}{\ls^2\over \ms}{\del \over
\del \ms}\right) V_{\MS}=0.
\eqnd\msrg$$
We want to emphasize that eq.~\msrg~above holds to all orders
in this scheme because mass divergences 
arise only at two loop order ($\lambda_2=\ms^2+{1\over 12(4\pi)^2}{\ls^2\over
\epsilon}$) and no additional finite renormalization is needed in the
$\MS$~scheme. In practice \msrg~should be satisfied at $b$ loops up to order
$O(\ls^{b})$ (if we take  $\phi\sim O({1/\sqrt{\ls}})$): 
the term  $\ls^2$ enforces relations between the coefficients of
$\log$, $\log^2$ at order $b$ and terms at order $b-2$.
The constant $F_{4,0,1}$ cannot be fixed because the corresponding relation 
implied by \msrg~involves terms without field dependence (i.e.~``cosmological
terms") that are not considered in \msveff.

Numerical results up to five loops for $F_{blk}$ 
have been first published in  Table II of 
\rBBMN~(derived from Table I that contains the values of
single diagrams, see also \rNMB). 
Some misprints in Table I, of eq.~(3.2) and Table II of \rBBMN~are reported
in \rHaDo, where Table II
is rederived  from the corrected values of Table I,
a corrected eq.~(3.2) and eq.~(B1--B3) of  \rBBMN.
We checked once more Table 1 of \rHaDo, and we obtained the same values,
provided that $-(-A_d)^b$ of eq.~(3.2) of \rBBMN~is replaced 
by $(-A_d)^b$ (a misprint is present in the prescriptions  of \rHaDo),
provided that
the formulae eqs.~(B1--B3) of \rBBMN~are intended to hold for the
dimensionless $\Gamma_b$ functions (i.e.~divided by a factor $g_0^3$
compared to definition eq.~(3.2)), and that ${\tilde X_0}$ is 
considered as dependent from ${\tilde r_0}$ when taking derivatives
${\del \over \del{\tilde r_0}}$ as in eq.~(B3). 
  
For our purposes we are interested in the physical mass scheme, that 
is fixed by normalization conditions  \egrzerom{}. The required  finite 
renormalization can be summarized by the following relation:
$$
V_M(\phi,m,g)=V_{\rm MS}( {\sqrt Z} \phi, Z_m m, Z_g m g,\bar\mu)\eqnd\ermstom
$$    
that defines the effective potential in the new scheme in terms of the
new renormalized mass $m$  and dimensionless coupling $g$.\par
The renormalization constant $Z$ is fixed by \egrzerom{} and we rewrite here
for completeness. It is obtained from the derivative of the
two-point function: 
$$Z^{-1}=\left({\del \Gamma^{(2)}_{\rm MS}\over \del p^2}
/\Gamma^{(2)}_{\rm MS}\right)_{p=0}=\left({\del \Gamma^{(2)}_0\over \del
p^2} /\Gamma^{(2)}_0\right)_{p=0}.$$
The last equality is due to the absence of field renormalization
between bare and $\MS$~quantities. Thus the
constant $Z$ is the same that one would obtain in passing from 
bare quantities to mass scheme and can be obtained from the 
known RG functions of the latter scheme:
$$ \log Z=\int_0^g \d g' {\eta (g')\over \beta(g')}. $$
The other constant can then be fixed by 
$$ \eqalign{&{d^2 V_M\over d\phi^2}(0,m,g)=m^2 \cr
&{d^4 V_M\over d\phi^4}(0,m,g)=m g\, .\cr }$$

Due to the fact that the potential in the mass scheme  will not depend
on $\bar\mu$, one can simplify tedious calculations by choosing
$\bar\mu\propto \ms$. 

The reduced potential $V$ is then obtained from:
$$
V(z,g)\equiv {m^3\over g} V_M(z\sqrt{m/g},m,g)
.$$
We report here the final (numerical) results for the coefficients 
$F_{l}$ of the Taylor expansion in $z$ of $F(z,g)$,
the  derivative with respect to $z$ of $V$ (see \eVpotFeq). 

%%%%%%%%%%%%%%F_L VALUES%%%%%%%%%%%%%%%%%%%%%%%%%%%%%%%%%%
$$
\eqalign{
{ F_5} &=+ {g\over {256\pi }} 
 -  {{{g^2}}\over {2048{{\pi }^2}}} + 
    {{{g^3}\left( -274 + 18{ k_3} - 24{ k_4} + 
          27\log ({3\over 4}) \right) }\over {442368{{\pi }^3}}}
\cr
& -2.25018021\times{{10}^{-7}}{g^4} + 
    1.977252\times{{10}^{-8}}{g^5}
\cr
{ F_7} &= - {g\over {1024\pi }}
 + 
    {{65{g^2}}\over {147456{{\pi }^2}}}  + 
    {{{g^3}\left( 646 - 40{ k_3} + 80{ k_4} - 
          45\log ({3\over 4}) \right) }\over {1769472{{\pi }^3}}}
\cr
&+4.17878698\times{{10}^{-7}}{g^4} - 
    4.512966\times{{10}^{-8}}{g^5}
\cr
{ F_9} &=  + {{5g}\over {16384\pi }} 
 - {{25{g^2}}\over {98304{{\pi }^2}}}
+ 
    {{{g^3}\left( -69099 + 4200{ k_3} - 11200{ k_4} + 
          3780\log ({3\over 4}) \right) }\over {339738624{{\pi }^3}}}
\cr &
-5.21924201\times{{10}^{-7}}{g^4} + 
    6.986358\times{{10}^{-8}}{g^5}
\cr
}
$$

%%%%%%%%%%%%%%%%%%%%%%%%%%%%%%%%%%%%%%%%%%%%%%%%%%%%%%%%%%%%
%%% notation a=4/i_4 /6 ==>i4/4=1/(6a)
\listrefs

\draftend
\end